\documentclass[aps, prd, twocolumn, groupedaddress, showpacs, showkeys]{revtex4}
\usepackage{graphicx}
\usepackage{dcolumn}
\usepackage{mathrsfs}
\usepackage{amsmath}
\usepackage{latexsym}

\newcommand{\ben}{\begin{equation}}
\newcommand{\een}{\end{equation}}
\newcommand{\bea}{\begin{eqnarray}}
\newcommand{\eea}{\end{eqnarray}}
\newcommand{\ba}{\begin{array}}
\newcommand{\ea}{\end{array}}
\newcommand{\bit}{\begin{itemize}}
\newcommand{\eit}{\end{itemize}}

\def\math{\mathsurround 0pt}
\def\oversim#1#2{\lower.5pt\vbox{\baselineskip0pt \lineskip-.5pt
        \ialign{$\math#1\hfil##\hfil$\crcr#2\crcr{\scriptstyle\sim}\crcr}}}

\begin{document}

\title{Oscillons and Domain Walls}
\author{Mark Hindmarsh$^{1}$}
\email{m.b.hindmarsh@sussex.ac.uk}
\author{Petja Salmi$^{1,2}$}
\email{salmi@lorentz.leidenuniv.nl}
\affiliation{$^{1}$Department of Physics \& Astronomy, 
University of Sussex, Brighton BN1 9QH, UK\\
                   $^{2}$Lorentz Institute of Theoretical Physics, 
University of Leiden, The Netherlands}
\date{\today}

\begin{abstract}
Oscillons, extremely long-lived localized oscillations of a scalar field, 
are shown to be produced by evolving domain wall networks in $\phi^4$ theory 
in two spatial dimensions.
We study the oscillons in frequency space using 
the classical spectral function at zero momentum, and 
obtain approximate information of their velocity distribution.
In order to gain some insight onto the dilute oscillon 'gas' produced by 
the domain walls, we prepare a denser gas by filling the simulation 
volume with oscillons boosted in random directions.
We finish the study by revisiting collisions between oscillons 
and between an oscillon and a domain wall, showing that 
in the latter case they can pass straight through with minimal distortion.

\end{abstract}

\keywords{Solitons, Oscillons, Breathers, Domain Walls}
\pacs{03.65.Pm, 11.10.-z, 11.27.+d}

\maketitle

\section{Introduction}

There is good understanding of the 
formation~\cite{Kibble:1976sj,Zurek:1985qw,Hindmarsh:2000kd} of 
coherent structures in phase transitions.
While the focus has been in time-independent solutions whose stability 
is protected e.g. by the conservation of a topological charge, 
there are time-dependent solitons as well, like Q-balls and {\it oscillons}. 
The stability of Q-balls~\cite{Coleman:1985ki} 
is under control by conservation of baryon number, but there is no 
evident guarantee for oscillons, and the longevity of these localised, 
non-perturbative oscillations is not well-understood.
They were found first already in 
the 70's~\cite{Bogolyubsky:1976nx,Bogolyubsky:1976sc} and then 
re-discovered~\cite{Gleiser:1993pt} when the 
dynamics of first order phase transitions and bubble nucleation was studied 
(for an extended investigation, see~\cite{Copeland:1995fq}).
{ It should be noted that oscillon is not the only term appearing 
in the literature for the phenomenon, 
but we adopt it hereafter for the rest of this study.}

Oscillons have attracted quite some attention recently.
A significant development has been the extension of oscillon solutions 
to gauge theories.
First discovery was made in gauged SU(2) model~\cite{Farhi:2005rz}, where 
(1+1)-dimensional simulations showed that with a Gaussian initial data 
the fields settle quickly into an oscillating state that has not seen to 
decay, when the scalar to vector masses in the theory are set 
to be $m_{H} = 2m_{W}$. 
More recently, this study has been extended to SU(2)$\times$U(1) model,
describing thus the complete bosonic sector of 
the Standard Model. 
The full three dimensional simulations reported 
in~\cite{Graham:2006vy,Graham:2007ds} point that the result 
holds also in the presence of photons and without the assumption of 
spherical symmetry.
Thus with the mentioned fine-tuned mass ratio there is an oscillon 
with energy of the order $10$~TeV, yet out of reach of the current particle 
accelators, 
but however much less massive than Q-balls (see e.g.~\cite{Dine:2003ax}). 
The importance here does not lie merely in extending the appearence 
of oscillons into a wider class of theories, but also
in the potential phenomenological consequencies.
Oscillons could namely provide the necessary non-equilibrium conditions 
needed for baryogenesis. However, one must bear in mind that an earlier 
investigation of the potential of thermal production of oscillons 
at electroweak scale came to a negative conclusion~\cite{Riotto:1995yy}.

Along the oscillon in the Standard Model, there are recent studies 
of oscillons in scalar 
theories~\cite{Fodor:2006zs,Saffin:2006yk,Arodz:2007jh}.
A dedicated examination in three dimensions was carried out
in~\cite{Fodor:2006zs} showing compelling evidence for a critical 
frequency minimising energy and the size of oscillon core.
Similar results were communicated in~\cite{Saffin:2006yk} keeping the 
dimension of the theory a free parameter oscillon 
lifetime dependence on the dimensionality was investigated 
(see also~\cite{Gleiser:2004an}).
Most recently, compact, non-radiating, periodic solutions 
in (1+1)-dimensional signum-Gordon 
model have been reported in~\cite{Arodz:2007jh}.

{As solutions of non-linear field 
equations, oscillons are well worth investigating, 
as are their effects if they are created in the Early Universe.
In order for oscillons to play a role, there is need for processes 
to initiate large oscillations in the field in different models.}
Oscillon formation has been reported after 
supersymmetric hybrid inflation~\cite{Broadhead:2005hn} as well as 
the QCD phase transition, where dense oscillating 
pseudosolitons in the axion field have been observed~\cite{Kolb:1993hw} 
(for a study of pseudo-breathers in sine-Gordon model 
see~\cite{Piette:1997hf}).
In~\cite{Kasuya:2002zs} oscillating field configurations were reported to 
form in close to quadratic potentials.
Recently~\cite{Gleiser:2007te} oscillons were found as a 
result of vortex-antivortex annihilation in two dimensional 
Abelian-Higgs model, providing thus another example of oscillons 
in gauge theory. 
{Once formed, oscillons could considerably influence the dynamics 
of the system as has been suggested for the case of 
the bubble nucleation process~\cite{Gleiser:2007ts}.}

While the oscillating energy concentrations have seen to form 
with a tiny initial density contrast in~\cite{Kasuya:2002zs}, 
the purpose of this study is to 
report oscillon formation via much more violent process, namely from the 
domain collapse in $\phi^4$ theory. For a related study 
see~\cite{Dymnikova:2000dy} where it has been shown that 
collision of two bubbles in first order phase transition can 
lead to formation of long-lived quasilumps.

The paper is organised as follows.
First we review shortly the basics of the classical 
spectral function and examine the 
signal of oscillons in spectral function. We go on simulations 
with random initial conditions, the formation of oscillons from 
collapsing domains. 
We tackle this also statistically using spectral function.
In order to have a good control of radiation we also 
prepare initial state of only oscillons moving into random directions and 
report the dynamics of this oscillon gas. We finish our study by reporting 
the off-axis collisions of oscillons and an oscillons and a stationary 
domain wall.

\section{Numerical Set-up}

\label{s:Models}

The Lagrangian for a single real scalar field~$\phi$ 
is given by
\begin{eqnarray}
  \mathscr{L} =
 \frac{1}{2} {\partial}_{\mu} \phi {\partial}^{\mu} \phi - V(\phi), 
 \label{lagrangian}
 \end{eqnarray}
and the equation of motion thus reads
\begin{eqnarray}
\ddot{\phi}-\nabla^{2}\phi + V'(\phi)=0. 
 \label{eqm}
\end{eqnarray}
In this study $V$ is the degenerate double-well quartic potential
\begin{eqnarray}
V(\phi)=\frac{1}{4} \lambda (\phi^2-\eta^2)^2.
 \label{quartic} 
\end{eqnarray}
The vacuum expectation value and couplings can be scaled out and set to 
unity. With that choice the minima are at $\phi = \pm 1$ 
and the local maximum at $\phi = 0$.

The field equation is evolved on a two-dimensional lattice 
with periodic boundary conditions using a leapfrog update and 
a three-point spatial Laplacian accurate to O($dx^2$).
The lattice spacing for the data shown is set to be $dx=0.25$ and 
the time step $dt=0.05$. We utilise periodic boundary conditions 
throughout this study because we wish to allow oscillons to move without 
hitting any boundaries as well as let domains grow in simulations of 
random initial conditions.


\section{Oscillons in the spectral function}

\subsection{Spectral function in the classical approximation}

The one-particle spectral function for a real scalar field~$\phi$ 
in the classical approximation was given in~\cite{Aarts:2001yx} by
\begin{eqnarray}
\rho \, (t,\boldsymbol{x}) = -\frac{1}{T}
\left\langle \, \Pi(t,\boldsymbol{x}) \, 
\phi(0,\boldsymbol{0}) \,\right\rangle, 
\label{spectral_function}
\end{eqnarray}
where $T$ is temperature and $\Pi$ the field momentum $\dot{\phi}$. 

The numerical implementation of the correlator in~(\ref{spectral_function}) 
is straigthforward in leapfrog discretization. In~\cite{Aarts:2001yx}
the following symmetrized definition was suggested
\begin{eqnarray}
\rho \, (t,\boldsymbol{x}) = -\frac{1}{T}
\left\langle \, \Pi(t + \frac{dt}{2},\boldsymbol{x}) \, 
\frac{1}{2} \left(\phi(0,\boldsymbol{0}) + \phi(dt,\boldsymbol{0}) \right) 
\,\right\rangle .
\label{spectral_function_discretized}
\end{eqnarray}

The classical spectral function at zero spatial 
momentum $\rho (t,\boldsymbol{p}=\boldsymbol{0})$ can be obtained 
from a volume average of~(\ref{spectral_function_discretized}). 
The spectral function in frequency space, $\rho(\omega,\boldsymbol{0}) 
\equiv \rho(\omega)$, can in turn be derived by performing Fourier transform.

We do not attempt to define temperature in what follows here, but 
merely adopt the correlator of the field and field momentum as a 
useful quantity to monitor in order to determine frequencies present in 
the system under the study. We simultaneously point out that our 
choice, inspired by spectral function, is not unique, but other correlators, 
like e.g. equal time 
correlator of $\phi$ and $\Pi$, could be utilised equally well.
The choice of the reference point $\boldsymbol{x}= \boldsymbol{0}$ of the 
field~$\phi$ in~(\ref{spectral_function}) obviously does not play a role in
a homegeneous system where no lattice site is in special position. 
This is not strictly true when e.g. a stationary oscillon is placed 
in the middle of a lattice. Instead of choosing one lattice site, 
simulations averaging over all points in the lattice were performed. 
However, even though homogeneity cannot be directly assumed, 
an approximation of the spectral function at zero momentum 
where the correlator in~(\ref{spectral_function}) is 
replaced by the product of average value of $\Pi$ at given time and 
$\phi$ at reference time $t=0$, 
$\rho(t) = \bar{\Pi}(t) \cdot \bar{\phi}(0)$, 
turns out to yield the same information in frequency space. 
Computationally this approximation is much more economical and will be 
utilised during the rest of this study.

\subsection{Oscillon Signal in the Spectral Function}

\begin{figure}
\begin{center}

\includegraphics[width=0.94\hsize]{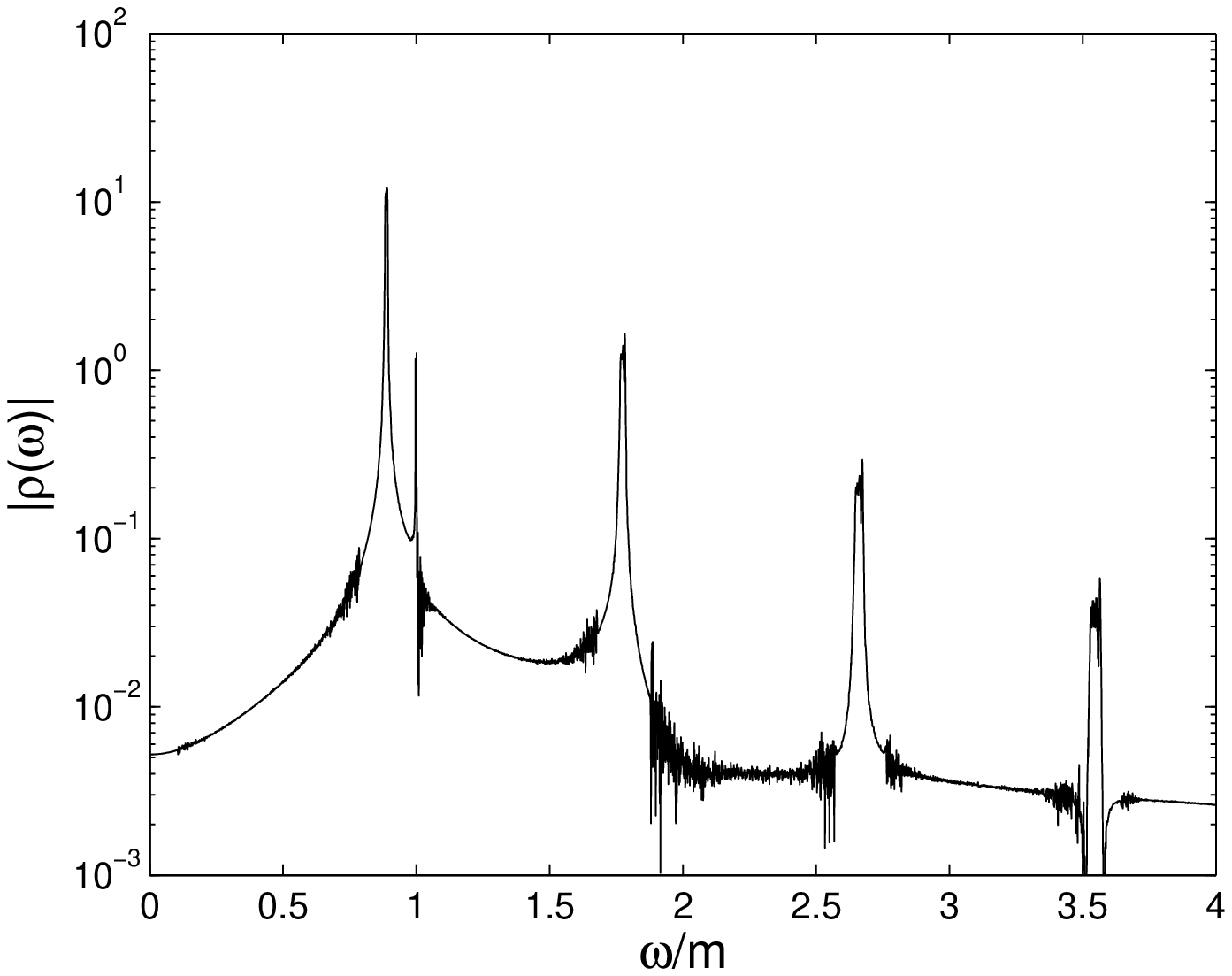}
\includegraphics[width=0.94\hsize]{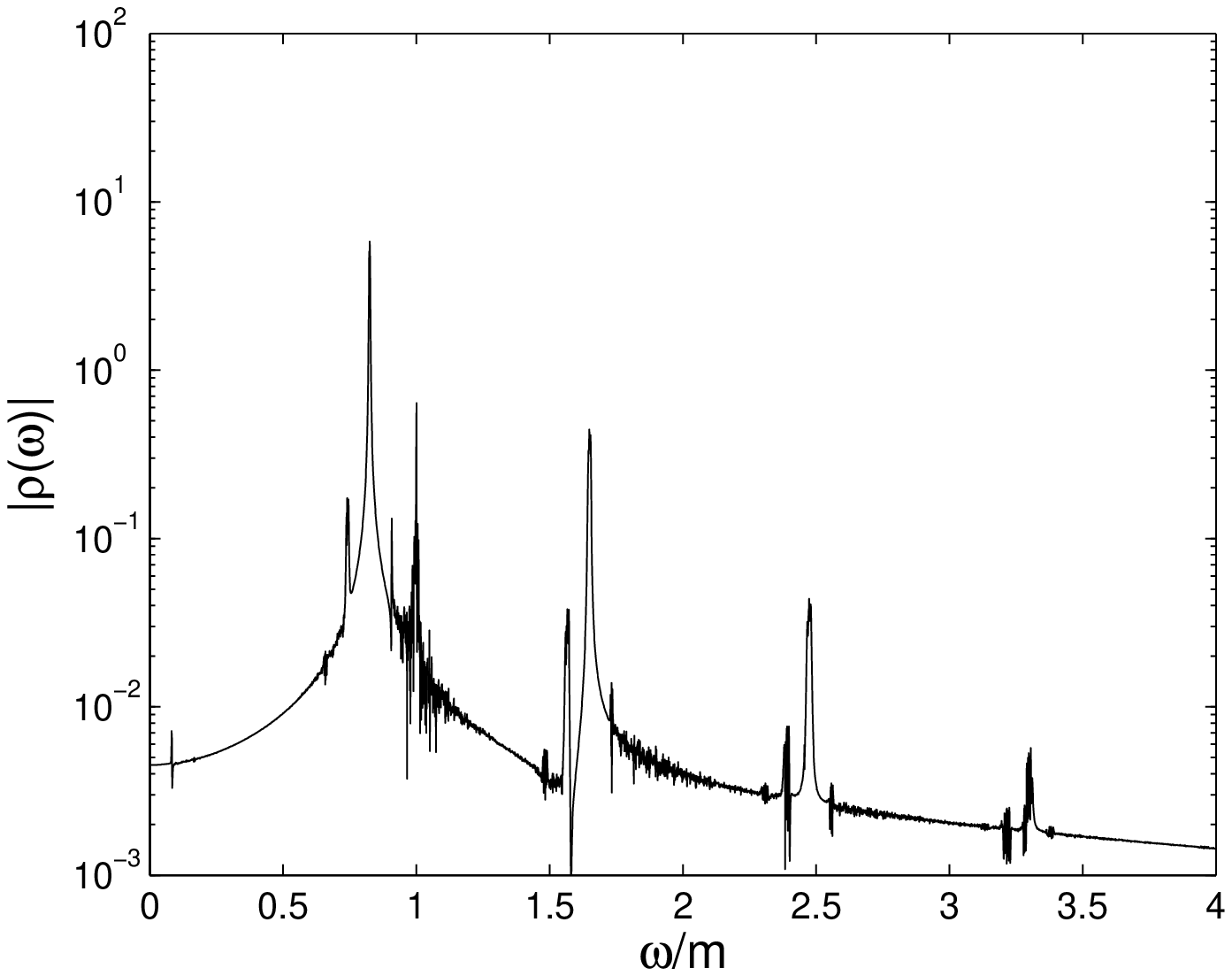}

  \caption{\label{f:spectral} The spectral function $|\rho(\omega)|$ at 
zero momentum over the time interval of length $5 \cdot 10 ^3$ for 
a stationary oscillon (above) and an oscillon with 
velocity $v \simeq 0.42$ (below).}

\end{center}
\end{figure}

\begin{figure}
\begin{center}

\includegraphics[width=0.94\hsize]{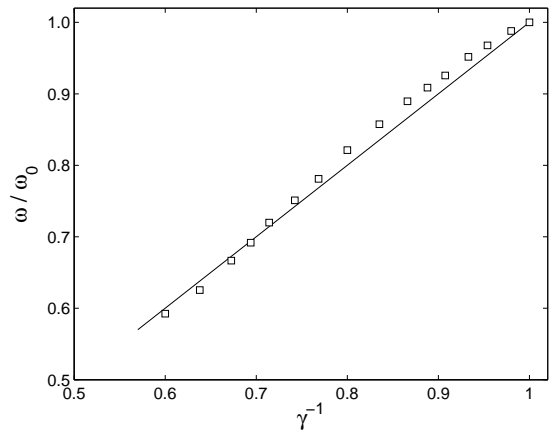}
  \caption{\label{f:frequency_vs_gamma} The relative 
frequency $\omega/\omega_{0}$ as a function of the inverse of 
the $\gamma$-factor. The solid line is $\omega=\omega_{0}/\gamma$.} 

\end{center}
\end{figure}

\begin{figure}
\begin{center}

\includegraphics[width=0.94\hsize]{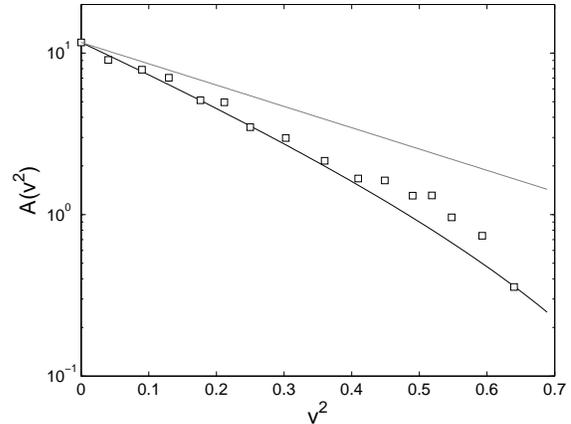}
  \caption{\label{f:amplitude_vs_velocity} The amplitude of the basic 
frequency, i.e. the power in the spectral function $\rho(\omega)$ at its peak, 
as a function of velocity. The black solid curve shows 
the right hand side of the Equation~(\ref{appendix4})
while the grey, straight line demonstrates the bare exponential 
$\exp(-\omega_{0}^2 r_{0}^2 v^2\, / \,2) $ without the additional 
power suppression.
The curves are fitted to have the value $11.63$ at $v=0$; 
$\omega_{0}=0.85$ and $r_{0}=2.9$.} 

\end{center}
\end{figure}

Oscillons are created on the lattice by using a Gaussian ansatz 
\begin{eqnarray}
\phi(r) =  \,( 1 - C \cdot \exp(-r^2/r_{0}^2)),
 \label{gaussian-ansatz}
\end{eqnarray}
where $r$ is the distance to the center of an oscillon 
$r=(x_{1}^2+x_{2}^2)^{1/2}$. The width of the distribution is set to be 
$r_0 \simeq 2.9$ (in units of  $(\sqrt{\lambda}\, \eta )^{-1}$), 
suggested an optimal choice in~\cite{Copeland:1995fq} and the 
maximum displacement $C=1$ so that the center of 
the oscillon starts from the local maximum of the potential. 
Oscillations take place at the basic frequency~$\omega_{0}$ that must 
be less than the threshold for radiation $m$, the mass in the theory.
The following periodic ansatz that also assumes spherical symmetry 
\begin{eqnarray}
\phi(r,t) = \sum_{n=0}^{\infty} f_{n}(r) \cos(n \omega_{0} t) \,,
 \label{solution-ansatz}
\end{eqnarray}
has seen to be rapidly coverging~\cite{Watkins,Honda:2001xg,Saffin:2006yk}, 
thus one expects also first few integer multiples of the 
frequency~$\omega_{0}$ to appear in the spectra. 

We perform a boost on oscillon and allow it then move on the lattice, 
the size of which for the data shown was set to be $400^2$. 
During the simulation we measure the mean $\bar{\Pi}(t)$ in order to 
attain $\rho(t)$. Figure~\ref{f:spectral} shows the 
spectral function $|\rho(\omega)|$ at zero momentum obtained by performing 
Fourier transform over the interval of length $5 \cdot 10^3$ in time 
units when an oscillon is stationary and when it is moving at velocity 
$v \simeq 0.42$. 
The location of the peak of the basic frequency of a moving oscillon 
is shifted left to a smaller value due to time dilation and its 
multiples correspondingly. 
The effect is best visible by comparing the location of the fourth peaks 
in the pictures or the distance between the peak of the basic frequency 
and that of radiation.
This drift is further illustrated in Figure~\ref{f:frequency_vs_gamma}, 
where the relative frequencies where the peaks appear in the spectra 
are shown against inverse of the $\gamma$-factor, $\gamma = 1/\sqrt{1-v^2}$.
The deviations of measured values from the straight 
line $\omega=\omega_{0} / \gamma$ illustrate the numerical limitations 
of determining the basic frequency in the data as well as the precision in 
the velocity of a moving oscillon.
The peaks at $\omega=m$ in Figure~\ref{f:spectral} indicate the 
presence of dispersive radiation component in the 
simulation box due to emission by oscillon.

In addition to the shift of the basic frequency there is much more 
drastic effect in the suppression of the height of the peak at the 
oscillation frequency as velocity of an oscillon increases. 
We expect exponential decrease (see Appendix). 
The value of the spectral function 
$\rho$ at $\omega=\omega_{0}\,(\gamma)$ is shown as a 
function of $v^2$ together with the prediction 
in Figure~\ref{f:amplitude_vs_velocity}.
We immediately point out that the precision in measuring the amplitude 
is not expected to be good. This is due to the restricted resolution
in the frequency (order $10^{-3}$ in units of $\omega/m$) 
when a discrete Fourier transform in a limited interval is carried out, 
whereas the peak itself is generally very narrow.
Consequently, the positions of the data points 
in Figure~\ref{f:amplitude_vs_velocity} can be considered only indicative. 
Alas, while time dilation shifts the location of the oscillation 
frequency further away from the radiation frequency thus easing the 
distinction between these two, the strength of the signal gets 
unfortunately strongly suppressed.

\section{Radiation of Oscillons from Collapsing Domains}

\begin{figure*}
\centering

\includegraphics[width=0.49\textwidth]{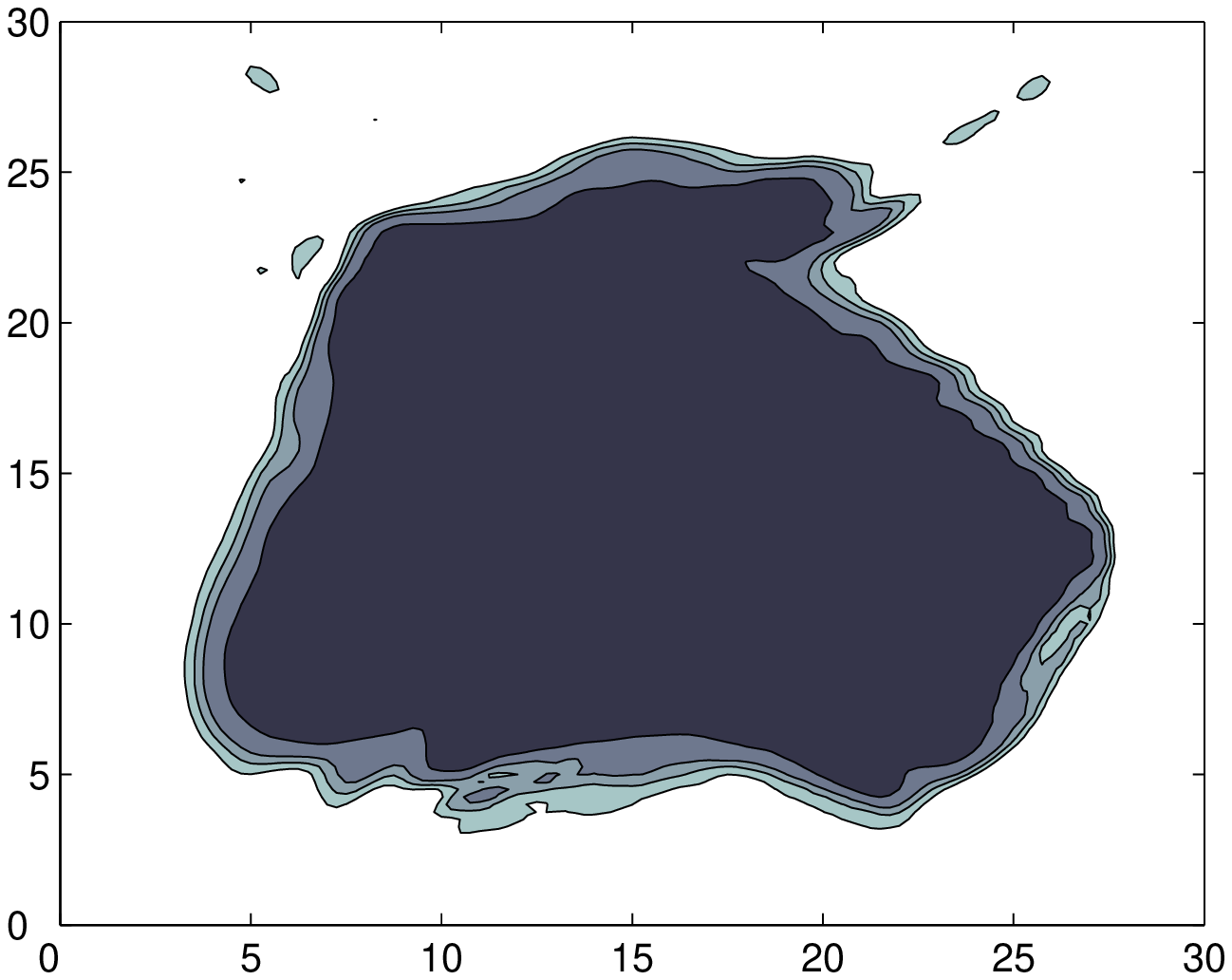}
\includegraphics[width=0.49\textwidth]{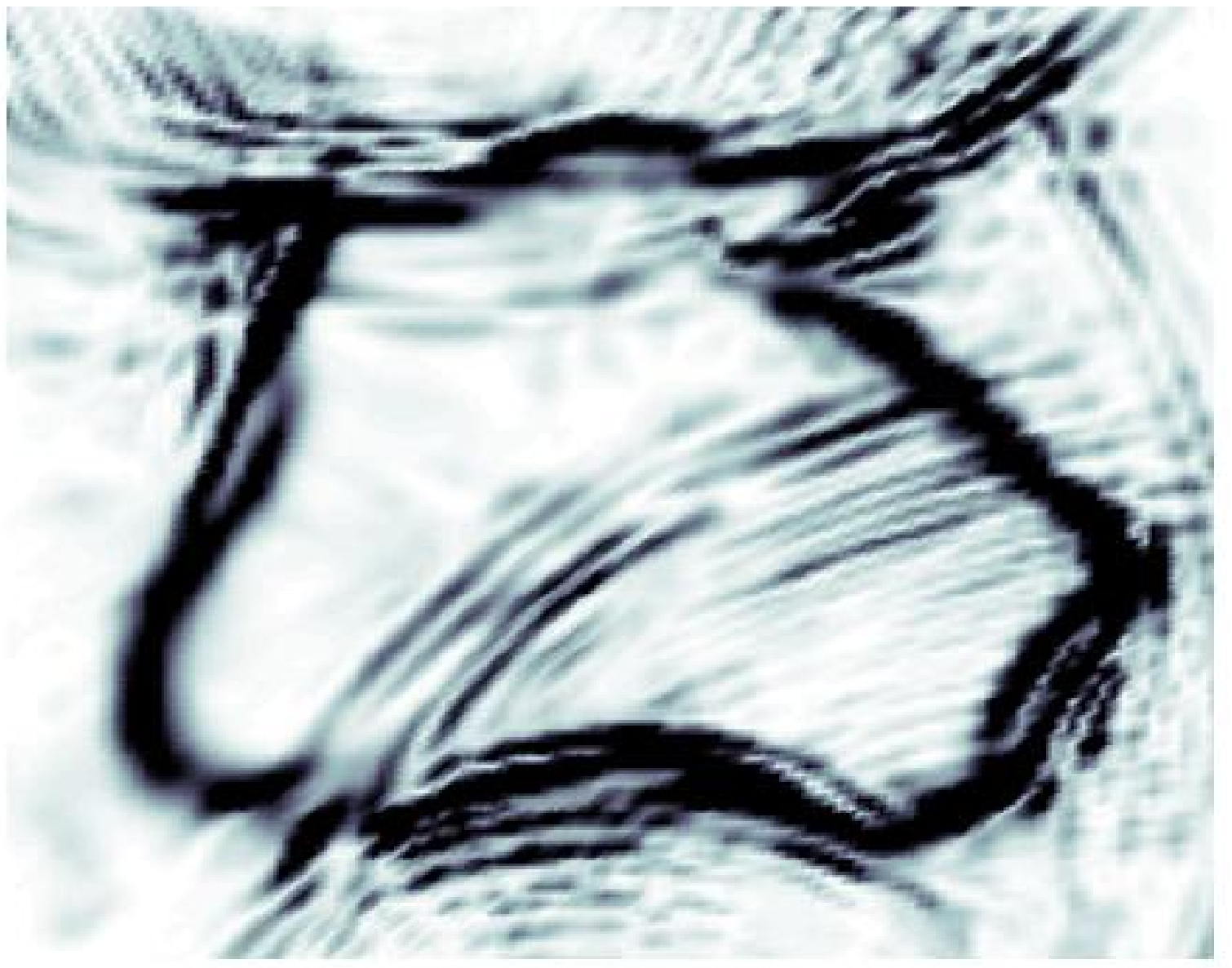}
\includegraphics[width=0.49\textwidth]{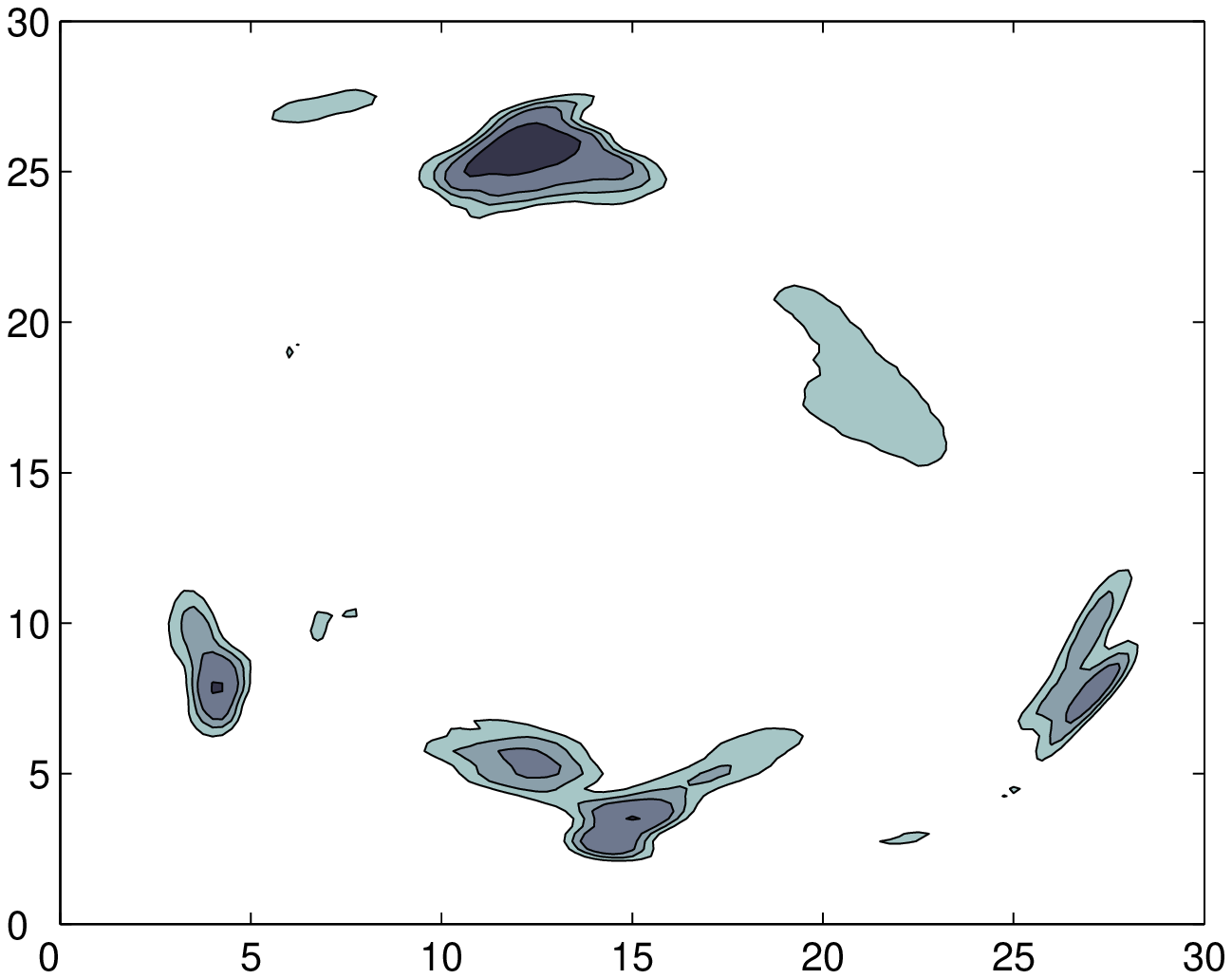}
\includegraphics[width=0.49\textwidth]{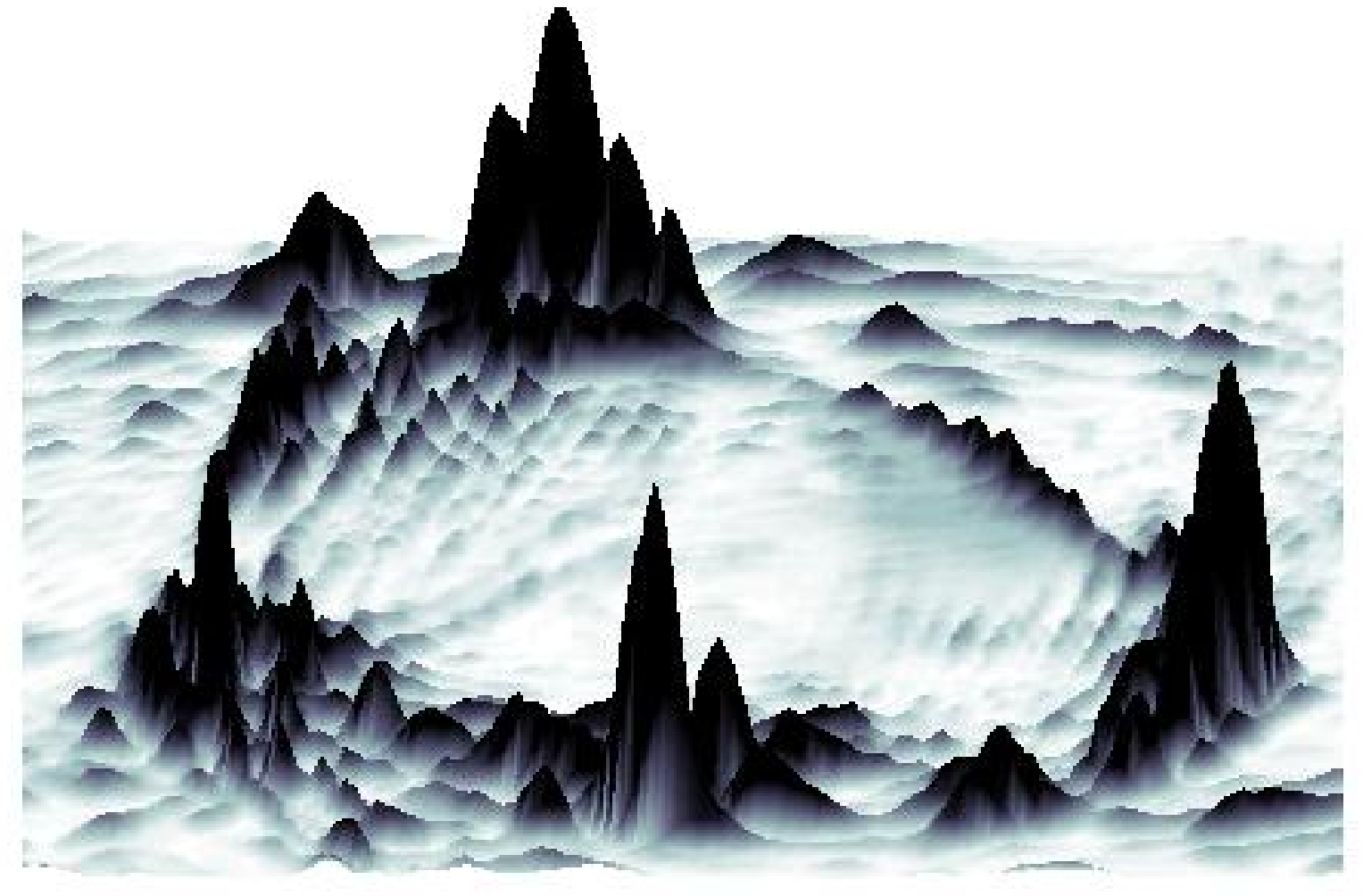}

  \caption{\label{f:domain_collapse} Oscillon formation in a 
domain collapse: the left panel shows contours of the field 
$\phi$ the right panel present the total energy density before the 
domain collapse (above) and right after (below). The snapshots 
are separated by $27$ time units; the physical size in linear dimension 
of the subbox shown is 30.}

\end{figure*}

\begin{figure}
\begin{center}

\includegraphics[width=0.94\hsize]{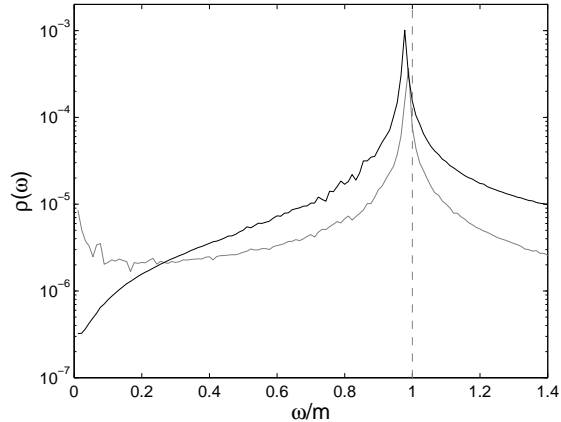}
  \caption{\label{f:spectral_random} The spectral function $|\rho(\omega)|$ at 
zero momentum over the time interval of length $400$ for oscillons (black) 
and domain walls (grey). The vertical dashed line shows the radiation 
frequency $\omega = m$ without fluctuations.}

\end{center}
\end{figure}

\begin{figure}
\begin{center}

\includegraphics[width=0.94\hsize]{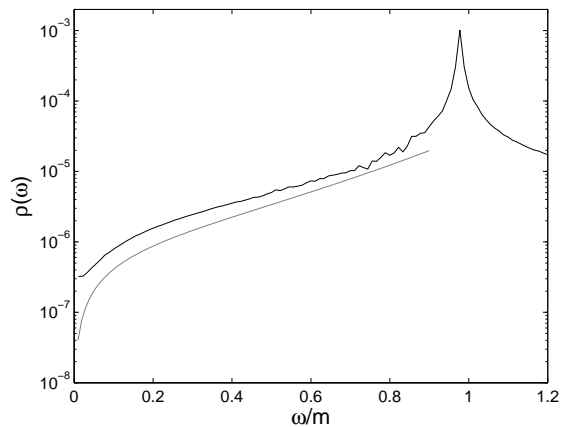}
  \caption{\label{f:spectral_fit} The spectral function $|\rho(\omega)|$ 
for oscillons (black) as in Fig.~\ref{f:spectral_random} together
with the curve $\omega\cdot \exp( b\, \omega^2)$ (grey), where $b=2.1$.}

\end{center}
\end{figure}

\begin{figure}
\begin{center}

\includegraphics[width=0.94\hsize]{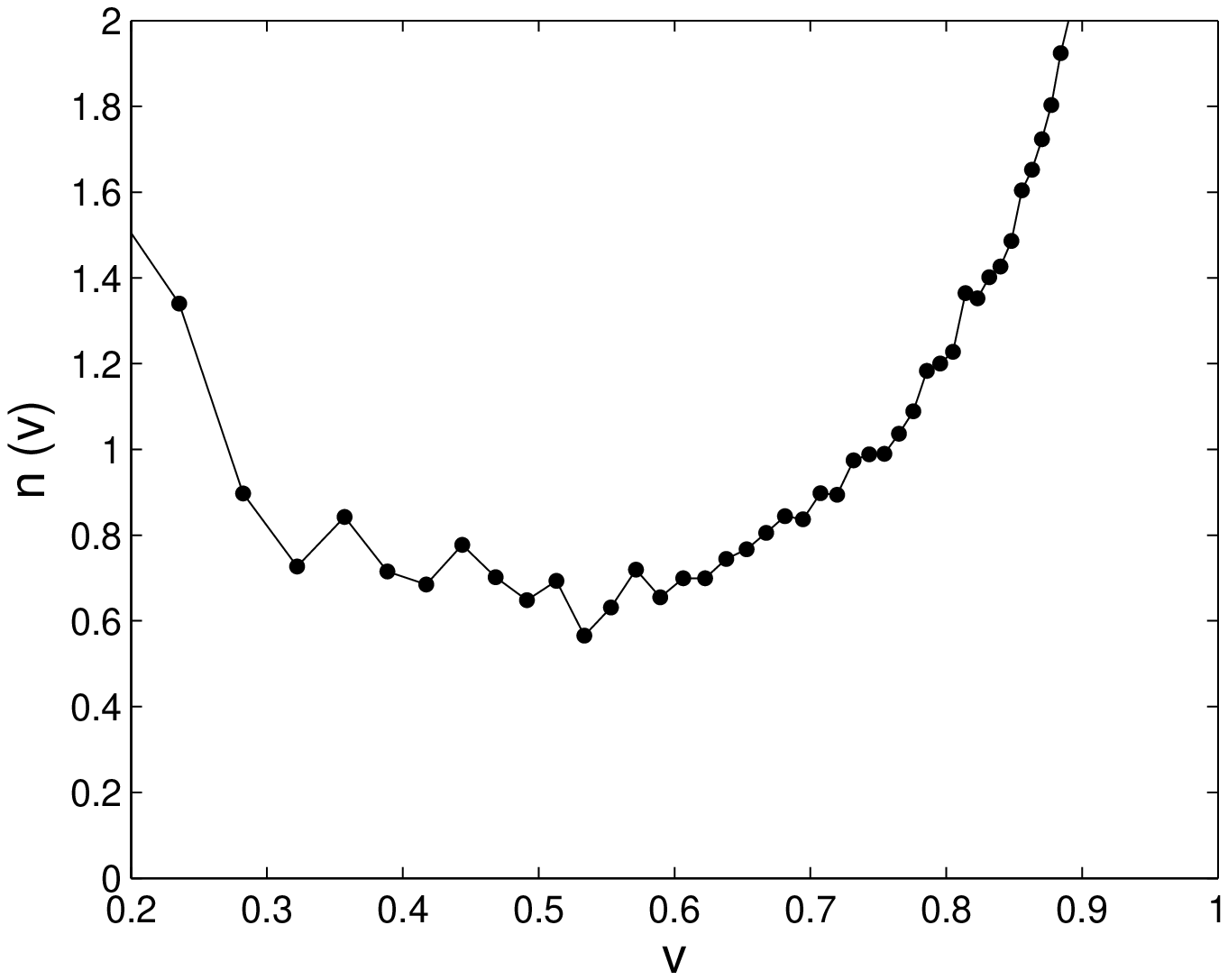}
  \caption{\label{f:velocity} The velocity distribution $n(v)$ derived 
from the signal in spectral function $\rho(\omega)$.}

\end{center}
\end{figure}

We evolve equation~(\ref{eqm}) numerically on a lattice of size $400^2$ 
(physical size in linear dimension $L$ is $100$) with random initial 
conditions. 
We use ``false vacuum'' initial conditions, the field is set to the 
local maximum of the potential and the field momentum is given a small 
random value picked from a Gaussian distribution with a zero mean. 
During the early stage the evolution is damped, but at time $t = 25$ the 
damping is turned off and then the field is allowed to evolve freely.
Domains where $\phi$ is in either of the two minima of the 
potential~(\ref{quartic}) are formed and separated by domain walls. 
As is well-known the domains grow in size during the evolution; 
a numerical study of the scaling properties of domain walls 
in classical $\phi^4$ theory is reported in~\cite{Garagounis:2002kt} 
and very recently confirmed in the quantum theory in the semiclassical 
approximation in~\cite{Borsanyi:2007wm}.

Obviously, a domain wall itself is an energy concentration due to the 
potential and gradient energy locked up in the wall, and a 
moving wall has naturally also a kinetic component.
When a domain collapses the energy that was trapped in the domain wall 
is released. Thus there is energy available for the field to 
execute large oscillations. 
The simulations carried out here point that usually the domain collapse takes 
place rapidly and then there tends to be large amount of 
energy, especially in the kinetic form, localised around the quickly 
shrinking domain wall already well before the eventual disappearance. 
In particular, there appear some very high 
energy concentrations moving along the wall. 
A similar kind of observation of energy concentrations, 
often (but not necessarily) along the domain wall, 
was made in~\cite{Copeland:2002ku} 
where field dynamics of tachyonic preheating after hybrid inflation 
was studied.
Once the domain has collapsed these energy concentrations 
either give rise to or turn into propagating non-linear waves. 
Oscillons are born.

The process of domain collapse is illustrated with two pair of snapshots.
Figure~\ref{f:domain_collapse} shows the isocontours of the field $\phi$ 
(left) and the energy density (right) before (above) and after (below) 
the domain collapse. Inside the contours dark grey indicates a region 
where the field is close to the disappearing vacuum, 
in the areas in lighter grey the field 
is around the maximum of the potential and white marks the field close to 
the other vacuum that becomes dominating one. 
In turn, the higher the energy density the darker the corresponding 
area is displayed, 
and the pentagon-like shape of the domain is also clearly recognisable in the 
corresponding snapshot of the energy density where the enclosing domain wall 
appears black.
In the second pair of snapshots the domain has collapsed - the large grey 
area has vanished indicating disappearence of this vacuum. 
Instead there are ripples where the field is far from the dominating vacuum 
positioned on a ring-like wave front.
These small, elliptic regions show up in extremely high peaks of 
energy density (the view in the picture is tilted 
to visualise the height of peaks of energy concentration).
These are oscillons which propagate along the spherical wave-front 
away from the 
location of the collapsed domain. Unlike the dispersive waves 
that are damped quickly, 
oscillons are far less dissipative and have a long range.
It should be noted that the elliptic form results from 
Lorentz contraction and thus indicates the high 
velocity of emitted oscillons. Furthermore,  
the seeds of oscillons are not required to be spherical: asymmetric bubbles 
can collapse into an oscillon~\cite{Adib:2002ff}.
The production of oscillons is associated with regions of high velocity on the 
domain walls, rather than bubble collapse where a 
single oscillon is formed from a bubble: there are generally several oscillons 
created by one collapsing domain. 

\subsection{Statistical Analysis in Frequency Space}

Eventually one or other of the two minima will completely dominate in a 
lattice of a finite size.
However, if the simulation box becomes divided between two domains 
that span over the whole volume, typically a fairly static, intermediate 
state follows where there are those two domains, 
and domain walls, present and which lasts for a long time.
We exploit this phenomenon in order to compare the signal of oscillons 
and domain walls in spectral function. 

It is relatively easy to observe if there are only one or two 
large domains present by monitoring the total length of domain walls 
in the lattice. If the total length of domain walls does not exceeds 
twice the linear size of the lattice $L$, there cannot be two domains 
spanning the box size. 
We determine the length of domain walls by counting the
number of lattice sites where the field changes sign compared to 
its nearest neighbouring lattice points. 
We work under the hypothesis that oscillons are created only 
when larger domains collapse, thus from the last domain of the disappearing 
vacuum and the ones from smaller domains created earlier disappear 
when colliding with the domain walls still present 
(though as demonstrated later on this is not inevitably the outcome). 

We generated $30,000$ different initial configurations. 
Since our attention is in the aftermath of collapsed domains, 
the simulations are evolved much longer than in a study interested 
in the scaling regime ($t \lesssim L/2$). 
We monitor the length of domain walls at two instants: $t=250$ and $t=750$.
Out of all the 
configurations, in 18,931 there is only one large 
domain dominating at the time of the first inspection.
In 8174 cases the simulation box was divided into two large domains 
throughout this interval up to the final time $t=750$.
In the remaining cases the collapse of the other domain takes place 
during this interval and these events are discarded here.

We examine the spectral function separately for these two above mentioned 
cases - single domain (oscillons) or two domains (and thus long domain 
walls present). 
Figure~\ref{f:spectral_random} shows the spectral function $|\rho(\omega)|$ 
obtained an average over all the selected configurations and 
normalised by the number of events for one domain present (black) and 
for two domains (grey). The Fourier transform is made 
over the time interval of length $400$ starting at time $t=350$.
This choice is to ensure that if there are shorter domain walls present 
at time $t=250$ that do not tricker the threshold $2L$ when 
monitoring is made, they have time to fade during the subsequent time 
interval before the observation interval commences.

For a single domain with oscillons the spectral function shows a fairly 
broad radiation peak, and a long ``shoulder'' of decreasing power together 
with almost complete absence of lowest frequencies.
This is in stark contrast with the growing power at small frequencies in 
presence of two domains. We believe this is the signal of the long 
domain walls that are very static objects thus contributing to the lowest 
frequencies in the system.
The peak at the radiation frequency is also narrower. This is easily 
understood by the fact that there is less kinetic energy in the system 
as large fraction of the energy remains locked up in the long domain walls. 
Thus these are cooler systems compared to those without domain walls. 
Not only the width, but also the location of the peak of radiation 
confirms this. In neither of the cases is the peak positioned 
precisely at $\omega = m$, but at sligthly lower frequency. 
This is because the effective mass in the theory is less than the bare 
parameter~$m$ due to the fluctuations of the field yielding a negative 
correction. In case of one domain the fluctuations are naturally larger 
and the shift is already visible effect yielding approximately 
2\% dislocation from $\omega = m$.
We doubt that temperature could be derived via kinetic energy in these 
cases yielding a sensible result 
because of the presence of oscillons which carry considerable kinetic 
energy, but should not be considered as thermal fluctuations.

The time dilation of the oscillation period relates the frequency and 
velocity via $\omega=\omega_{0} / \gamma$ and it 
is thus appealing to try to extract information of the velocity of 
oscillons from the power in the spectral function 
assuming that the signal in spectral function at 
frequencies~$\omega \lesssim 0.85$ is mainly due to the oscillons.
Figure~\ref{f:spectral_fit} shows that the "shoulder" can be reproduced.
The signal in the spectral function $\rho(\omega)$ can be transferred to 
an arbitrary velocity distribution $n(v)$ with the use of the 
amplitude given by~(\ref{appendix4}). One now needs to assume that 
the formed oscillons have typically size $r_{0} \approx 3$ - 
this is supported by the simulations.
Figure~\ref{f:velocity} 
shows the velocity distribution derived by~(\ref{appendix6}) from the 
signal in spectral function. The result is noisy, 
the formula~(\ref{appendix6}) diverges when $\omega \rightarrow 0$ 
and it cannot be expected to yield correct result at higher velocities. 
Unfortunately, the relation $\omega=\omega_{0} / \gamma$
also transforms the discrete signal in frequency space to the least number 
of data points at the low velocity end of the distribution $n(v)$.
Rather the result suggests that low velocities are dominant. 
This is presumably true: the velocity distribution is not measured at time 
of formation, but considerably later on when oscillons have lost 
energy in the strongly radiative environment 
(oscillon survival in thermal environment have 
been studied in~\cite{Gleiser:1996jb}).

\section{Oscillon Gas}

\begin{figure}
\begin{center}

\includegraphics[width=0.94\hsize]{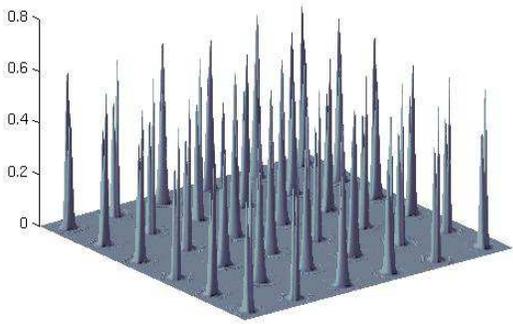}
\includegraphics[width=0.94\hsize]{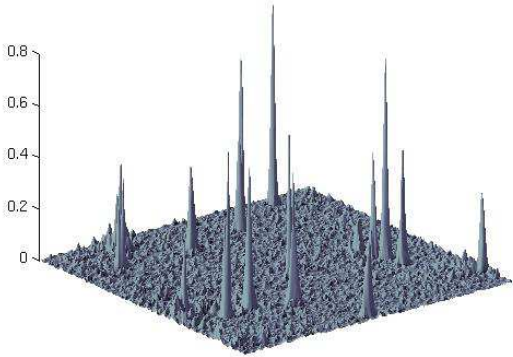}
\includegraphics[width=0.94\hsize]{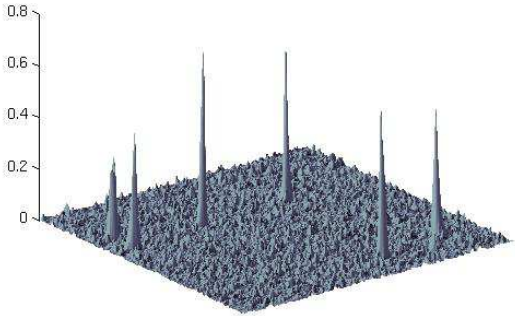}

  \caption{\label{f:gas} Sequence of snapshots showing the total 
energy density at times $t=0$, $937.5$, $10190$.}

\end{center}
\end{figure}

\begin{figure}
\begin{center}

\includegraphics[width=0.94\hsize]{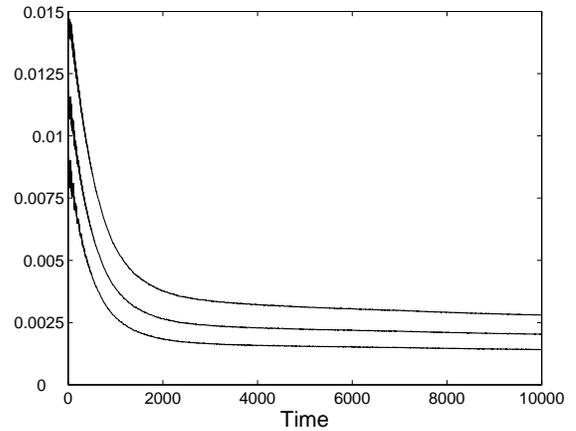}
  \caption{\label{f:gas_lin} Fraction of points of the lattice 
where the total energy density exceeds $0.15$, $0.2$ or $0.25$ 
($20$, $28$ and $35$ times the average from top to bottom) 
as a function of time.}

\end{center}
\end{figure}

\begin{figure}
\begin{center}

\includegraphics[width=0.94\hsize]{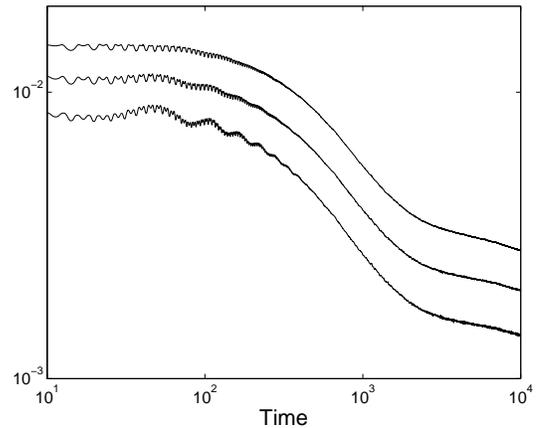}
  \caption{\label{f:gas_log} The same data as in Figure~\ref{f:gas_lin} 
on a logarithmic scale.}

\end{center}
\end{figure}

While simulations with random initial conditions and the subsequent formation 
of oscillons from collapsing domains provide reasonably realistic model of 
how oscillons could be produced e.g. in conditions present in the Early 
Universe, the drawback is the strong radiative component.
Inevitably, large fraction of the energy of the domain walls is realesed 
into dispersive modes that not only appear as the dominating signal in the 
spectral function, but also undoubtedly effect significantly the evolution 
of oscillons as discussed earlier. 

In order to simulate the evolution of an ensemble of oscillons in a far 
less radiative environment, we followed a different approach. 
An {\it oscillon gas} is initialised by preparing an oscillon lattice 
where oscillons are placed initially at equal distance from each others, but 
boosted with a velocity $v$ into random directions. In the subsequent 
evolution oscillons collide resulting to scattering, decaying and 
merging. The energy released from oscillons creates a radiative component, 
but this is starkly suppressed compared to the one present in simulations 
reported in the previous section.

Figure~\ref{f:gas} shows three snapshots of the total energy density in a 
simulation where initially 36 oscillons at velocity $v=0.5$ into random 
directions are placed on the lattice of size $800^2$ (physical size in linear 
dimension is $200$). There is a reduction in the number of 
oscillons and energy in dispersive propagating waves is clearly 
visible in the two later snapshots, but this is far less strong than 
those originating from domain walls.
Most importantly, the simulations show that few oscillons, typically 
half a dozen, survive for a long period of time, far greater than 
e.g. the time required an oscillon at initial velocity $v$ 
to travel around the lattice.

Because radiation is suppressed, higher concentrations in the energy density 
can be used to track oscillons. The survival of oscillons was studied 
more quantitatively by performing simulations of a larger ensemble. 
One hundred different initial states each having 121 oscillons at the 
beginning were evolved on a lattice of size $1500^2$.
Figure~\ref{f:gas_lin} shows the fraction of lattice points where the 
total energy exceeds $0.25$, $0.2$ or $0.15$ as a function of time 
(these thresholds correspond roughly 35, 28 and 20 times the mean value, 
respectively). For clarity the data is averaged over one oscillation 
period (roughly $5$ in time units). This is because all the oscillons start 
initially at the same phase and there is considerable variation in the 
hight and width of the energy density within the 
period (see~\cite{Hindmarsh:2006ur}). Some oscillations are still visible 
left in Figure~\ref{f:gas_log}, where the same data is plotted on a 
logarithmic scale. 
All the curves yield similar time evolution thus we conclude that there is no 
sensitivity to the thresholds chosen. As the dispersive waves do not 
contain such energy concentrations that would exceeds these thresholds 
as can be seen in Figure~\ref{f:gas} we further argue that the signal 
must be due to oscillons.

The data in Figures~\ref{f:gas_lin} and~\ref{f:gas_log} clearly show that 
there is decrease in number of lattice sites where energy density exceeds 
the thresholds corresponding demise of oscillons in collisions. However, 
the main result is that there are two phases: 
steeply declining slope flattens around time $t = 10^3$ to a 
much less rapid decay. At the end of the simulation there is 
still almost $20$~\% left of the initial 
number of lattice points where the thresholds are exceeded.

{In the region of fast decline the slopes have values 
approximately $-0.7$. This is marginally consistent with a decay law 
inversely proportional to time. Such a power law is be predicted by 
a simple annihilation picture: oscillons have constant velocity $v$ 
and cross section $\sigma$. Then in two dimensions their number 
density $N = N(t)$ obeys the differential equation 
\begin{eqnarray}
\dot{N} = - \left\langle v \sigma \right\rangle N^2,
\label{diffeq}
\end{eqnarray}
which yields time dependece $N \sim t^{-1}$.

While the steeper slopes can be understood on the basis of the 
differential equation~(\ref{diffeq}), we do not have 
any quantitative explanation for the cross-over.
On the qualitative level} there are at least two effects that could 
lengthen the life time of oscillons at a later stage 
of the simulations.
Firstly, collisions between oscillons, though not necessarily 
leading to a demise, cause considerable perturbation and 
oscillon radiates strongly before 
it settles back into a long-living state. If there occurs another collision 
during this relaxation stage that is then presumably highly likely to 
destroy the already perturbed oscillon.
In simulation it has been withnessed that the second collision 
indeed is often the fateful one.
Lower number density and consequently a longer mean free path 
can yield an enhancement of the survival probability in collisions. 
More importantly, due to the interactions oscillons slow down drastically 
(a decrease in the average velocity 
$\left \langle v \right\rangle$ can be partially responsible of 
the flatter slope than the expected $-1$ at the earlier stage). 
The reduced collision rate in turn increases the lifetime.

\section{Oscillon Collisions - Merging and Scattering}

\begin{figure*}
\centering

\includegraphics[width=0.41\textwidth]{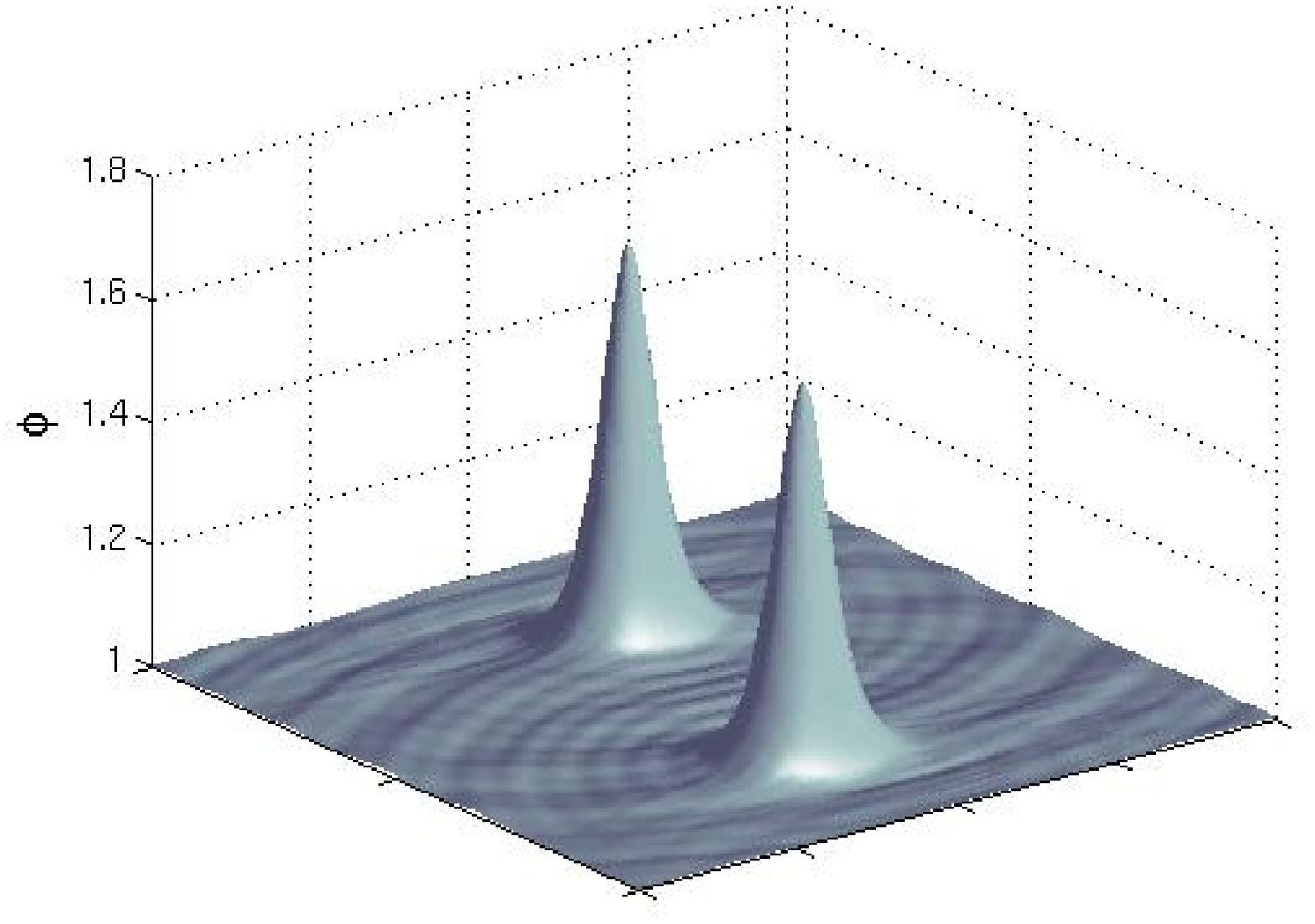}
\includegraphics[width=0.41\textwidth]{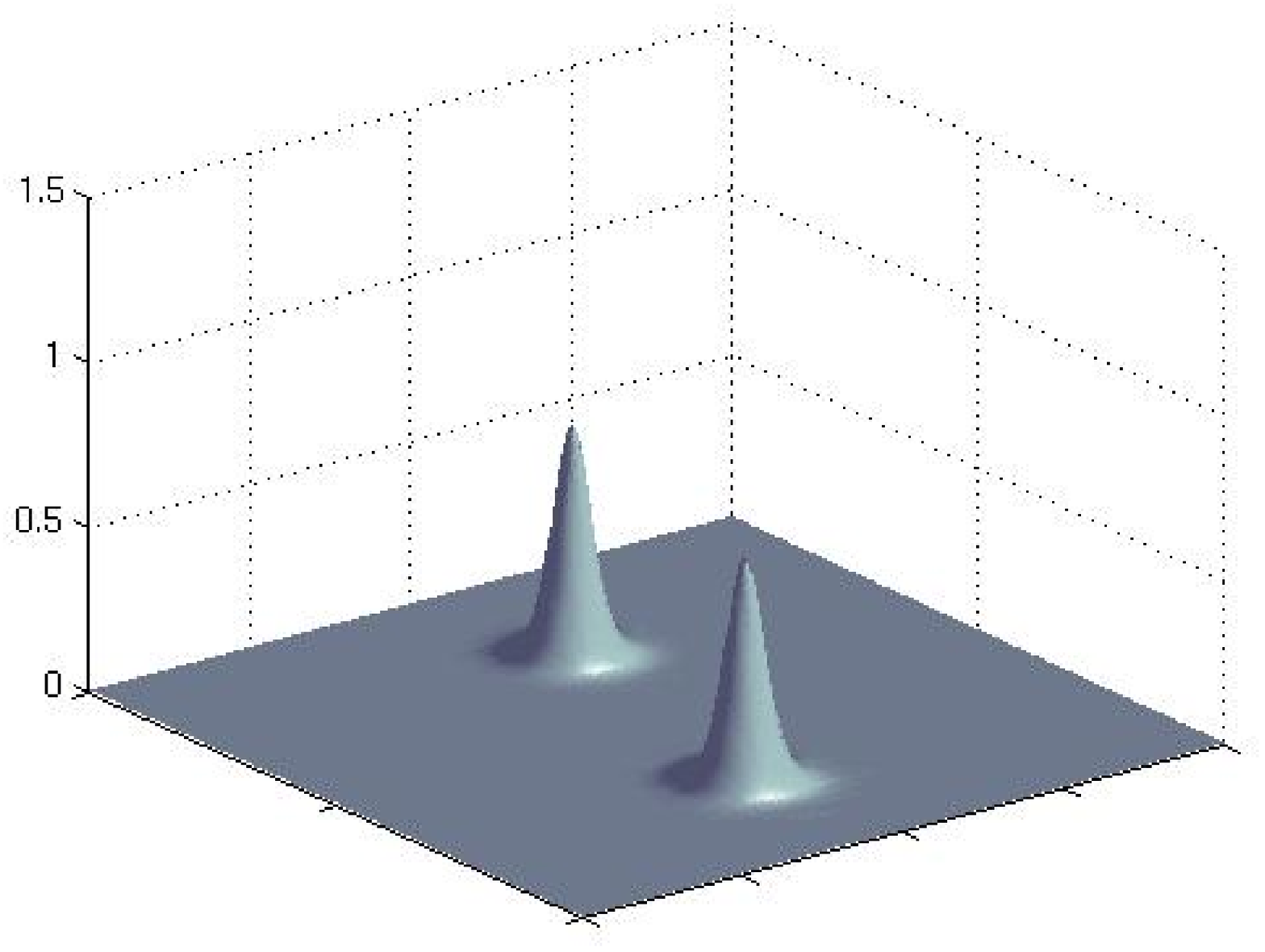}
\includegraphics[width=0.41\textwidth]{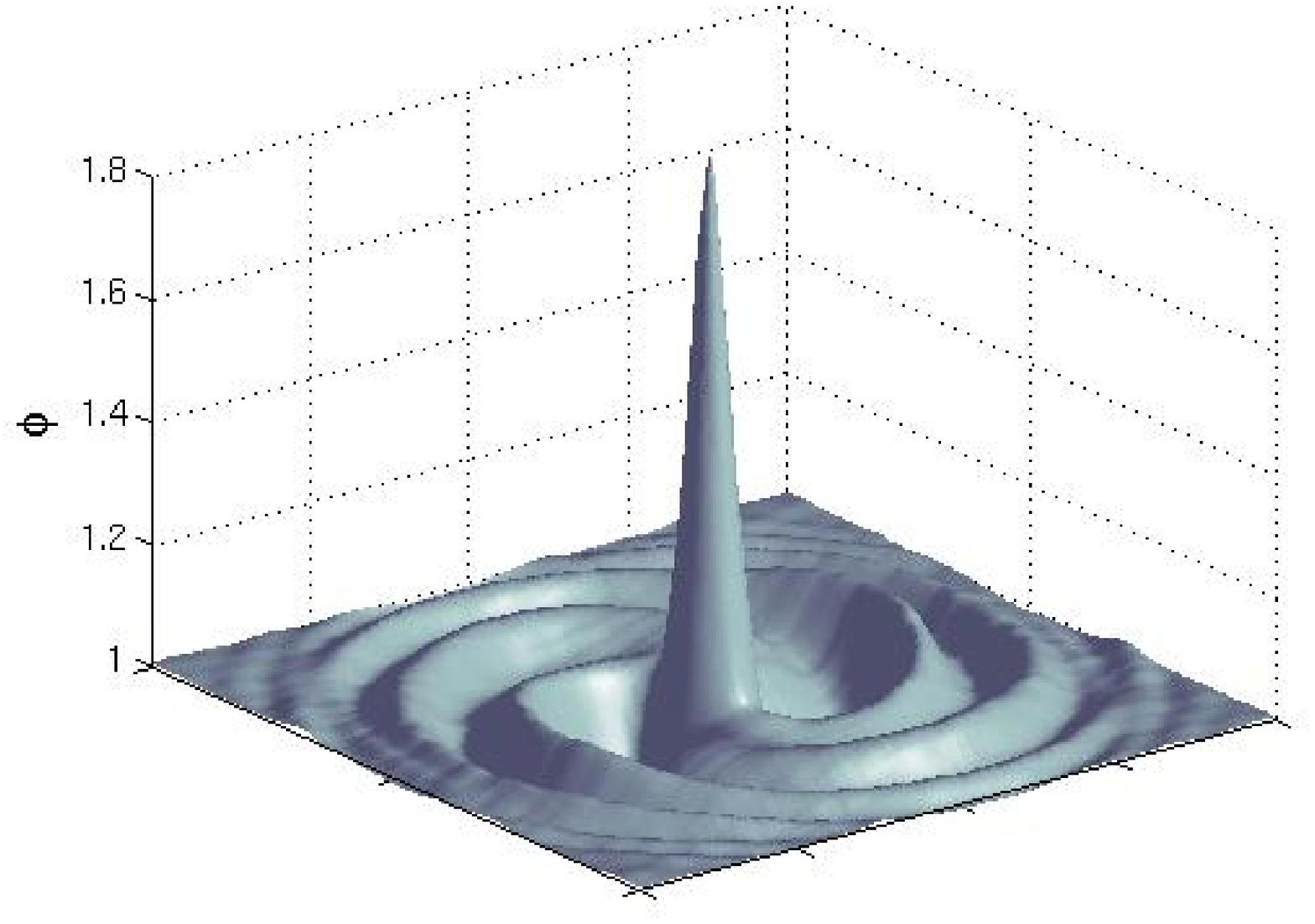}
\includegraphics[width=0.41\textwidth]{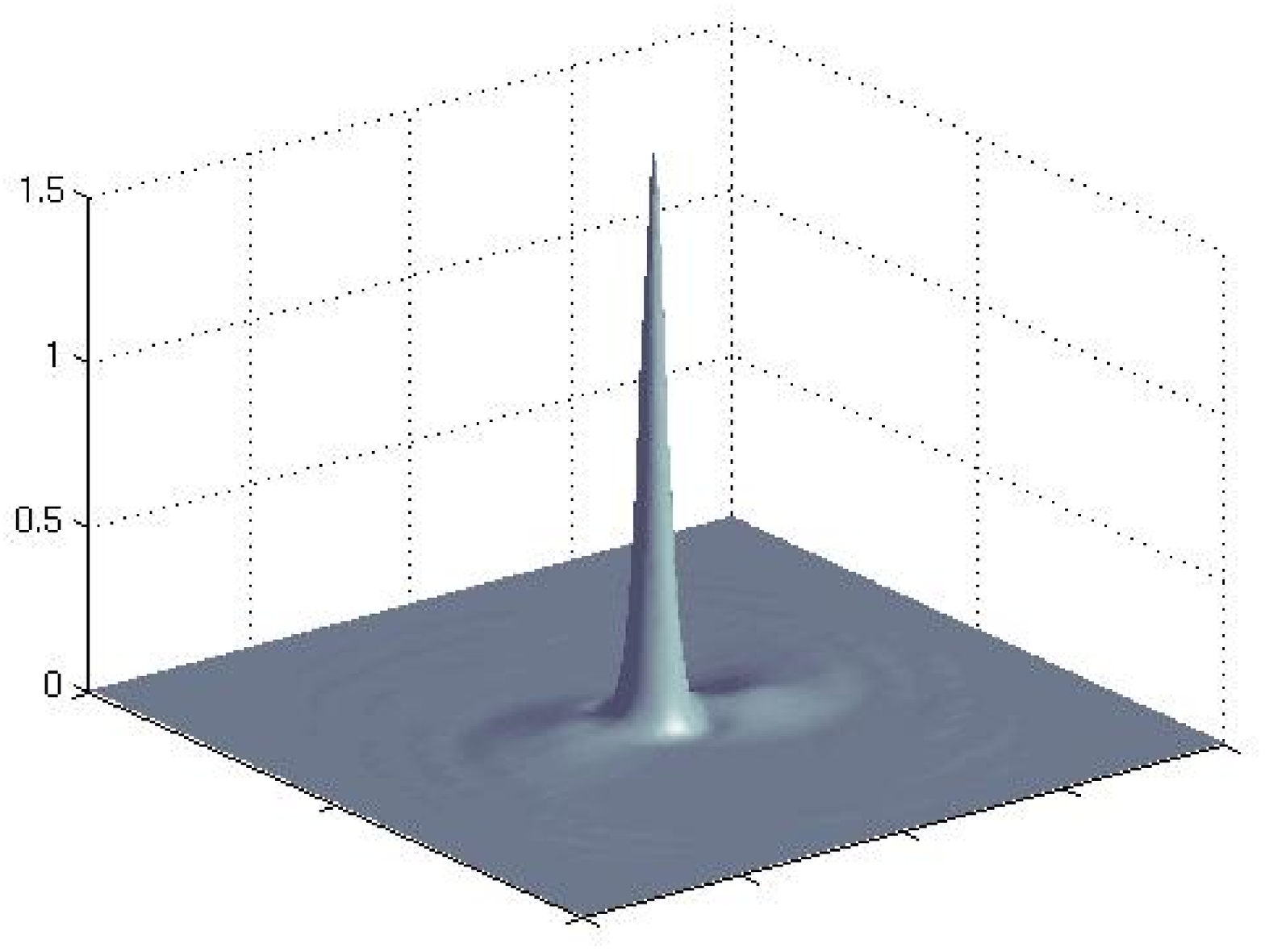}
\includegraphics[width=0.41\textwidth]{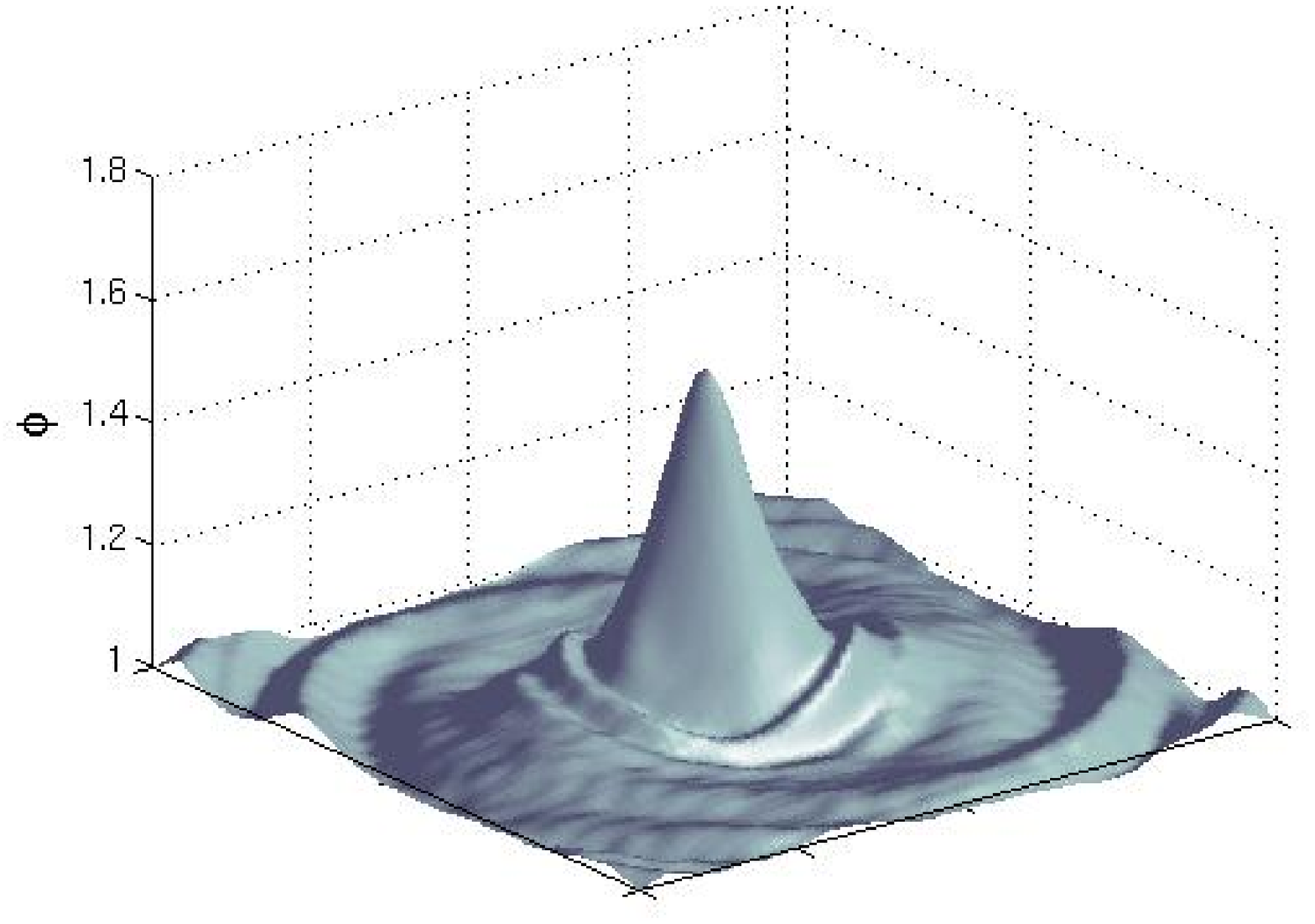}
\includegraphics[width=0.41\textwidth]{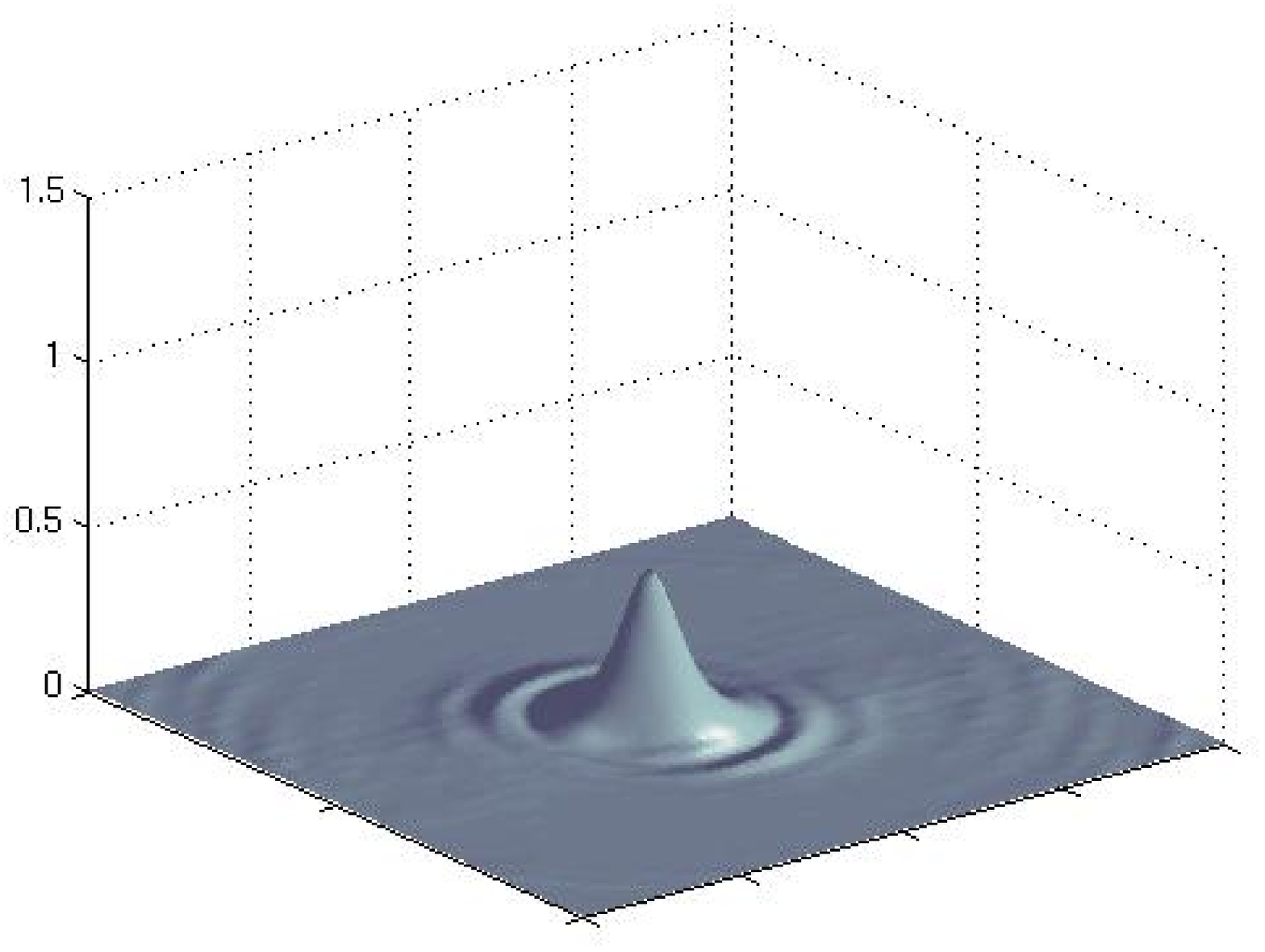}

  \caption{\label{f:merging} 
Oscillons merging: the field (left) and the energy density (right) at three 
instants (chronologically from top to bottom) in an off-axis 
collision of oscillons at velocity $v \simeq 0.2$. 
Alignment of the centers is displaced by $5.0$ units.}

\end{figure*}

The simulations presented in previous section show that collisions 
between oscillons can, apart from scattering or demise, 
result to merging. Off-axis collisions of oscillons were briefly 
studied in~\cite{Hindmarsh:2006ur} reporting an attractive 
scattering. Reducing the velocity causes the trajectories of oscillons to 
bend more after the encounter. 
There is a critical velocity, below which the oscillons do 
not scatter anymore, but merge together.
Figure~\ref{f:merging} shows snapshots of the value of the field as well 
as the energy density in an off-axis collision of two oscillons with 
velocity $v \simeq 0.2$ when the displacement in the alignment 
between the centers of oscillons, the impact parameter, is $5.0$ units. 
Considerable amount of energy is leaked in the process as can be 
seen from the spiral waves in the field. However, the energy density is 
simultaneously still concentrated in a very narrow region. 
Eventually, the resulting oscillon 
is deformed and the energy density is spread to a larger area.

\section{Oscillons crossing the domain walls}

\begin{figure*}
\centering

\includegraphics[width=0.41\textwidth]{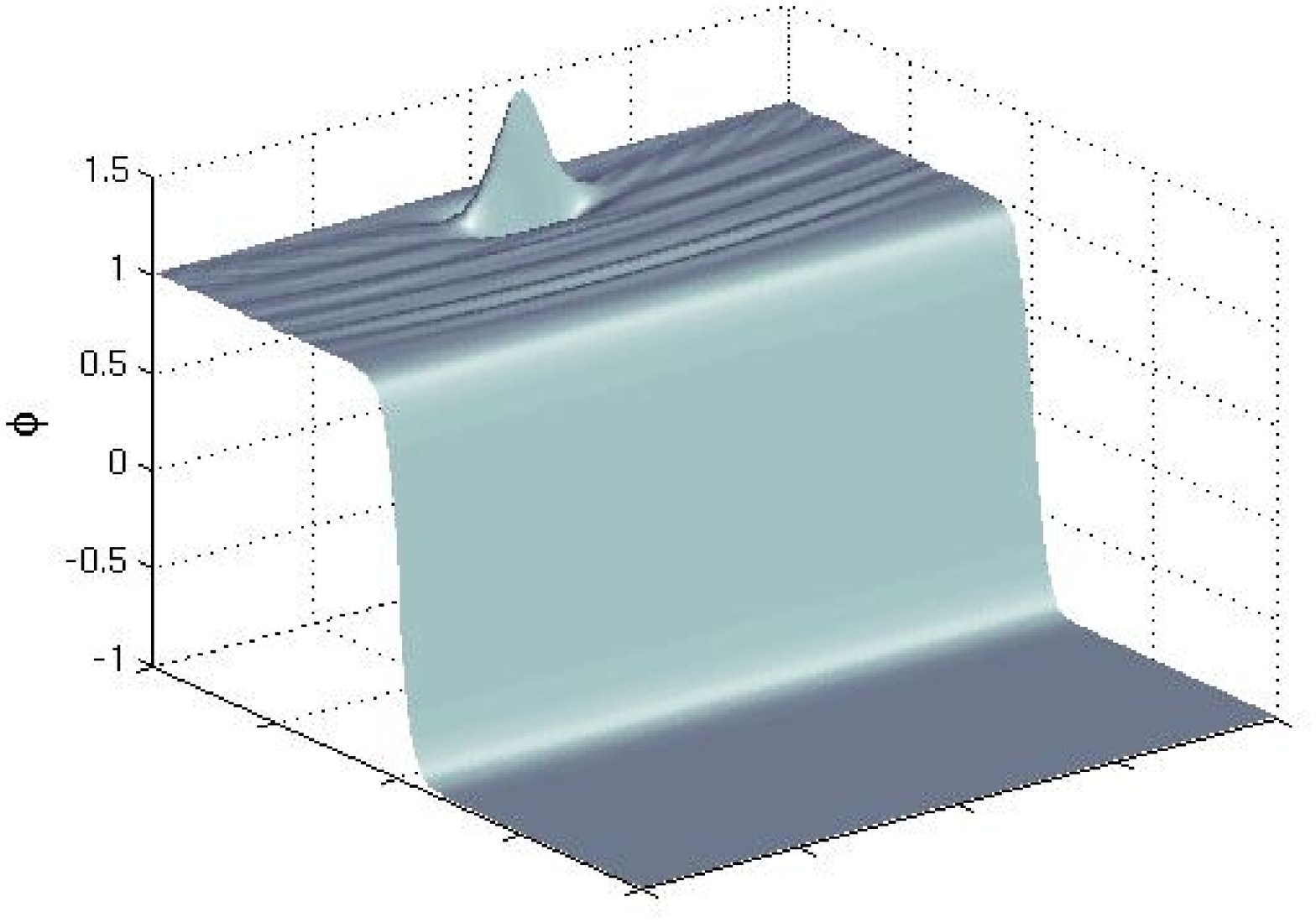}
\includegraphics[width=0.41\textwidth]{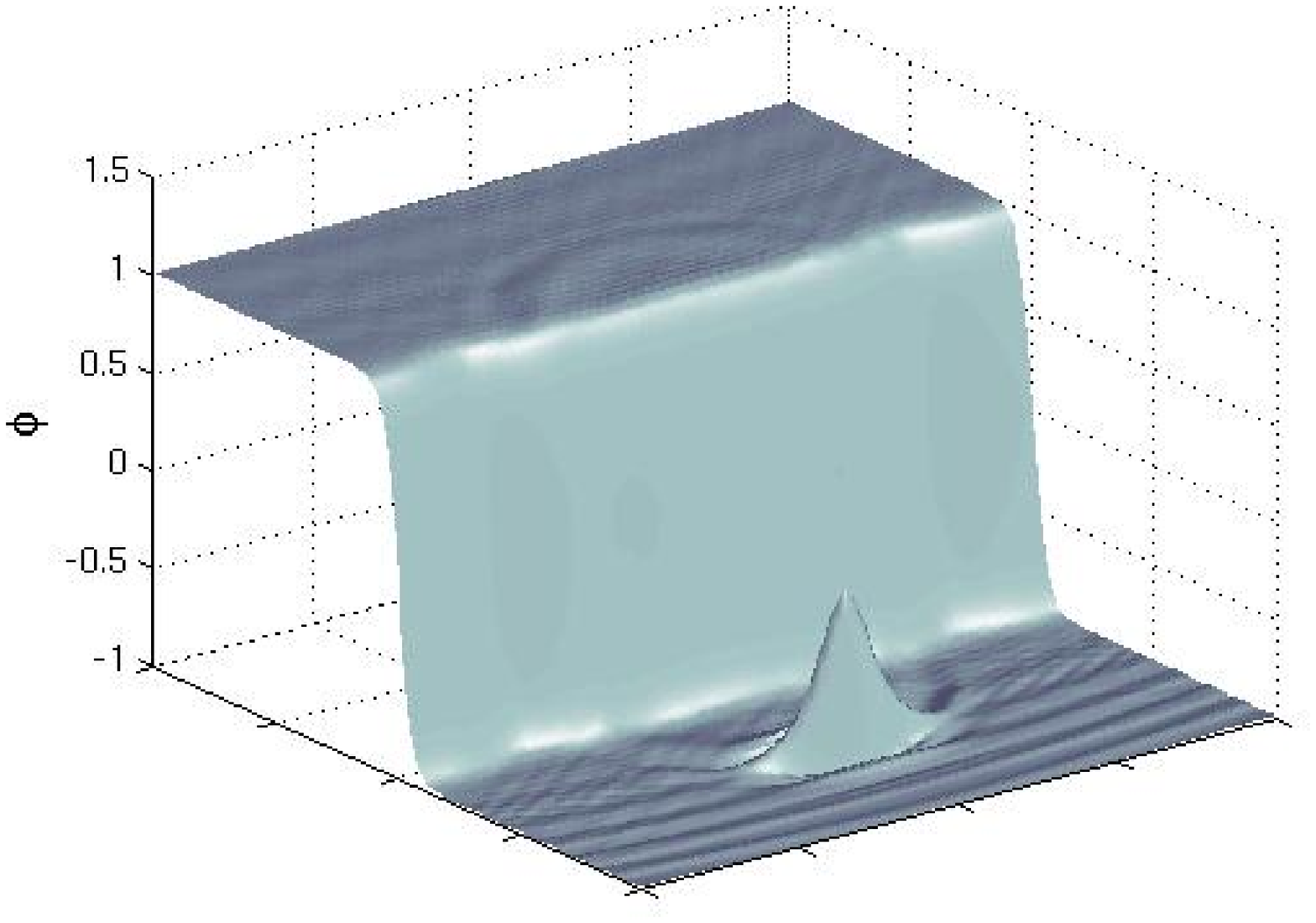}
\includegraphics[width=0.41\textwidth]{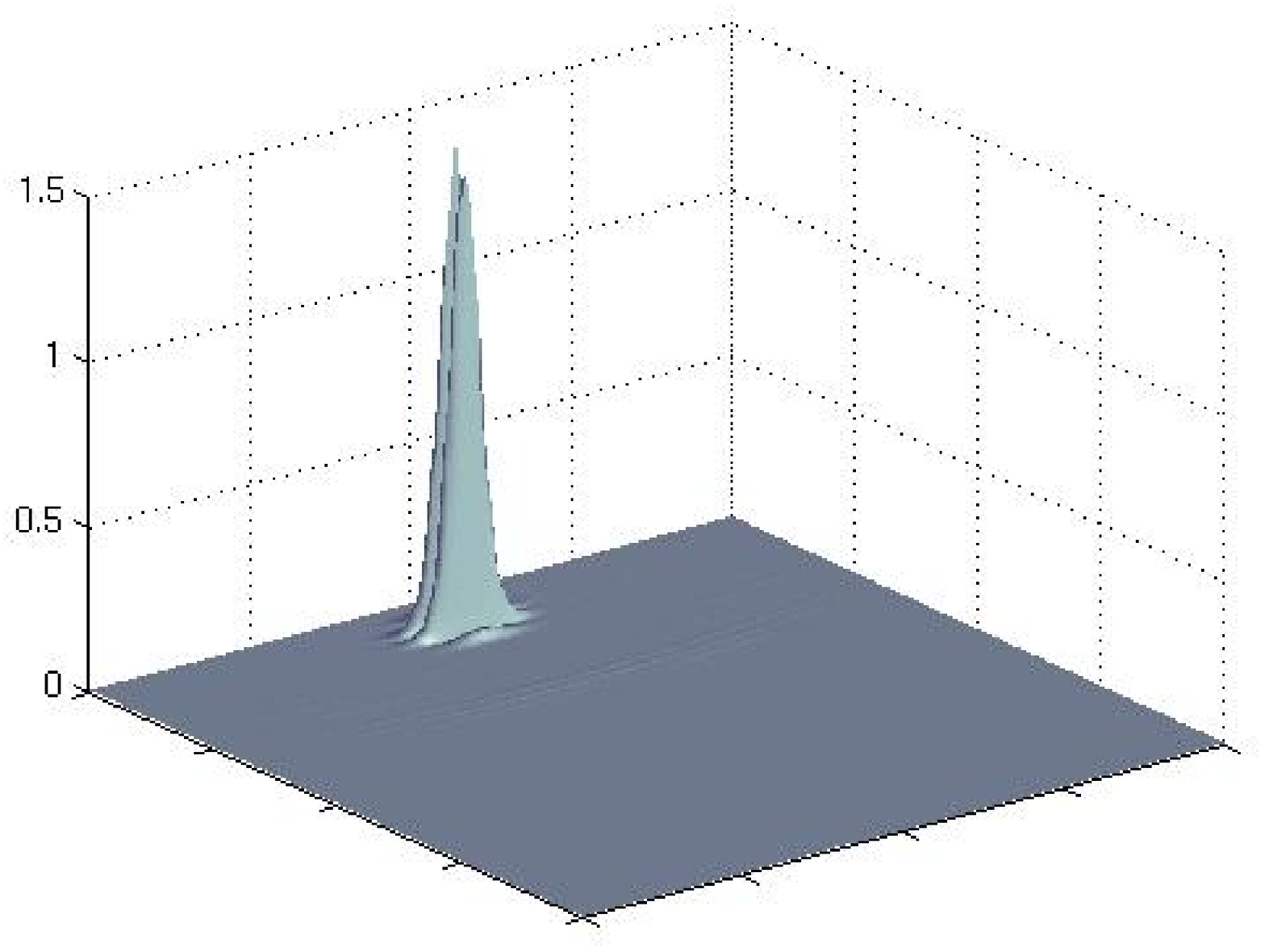}
\includegraphics[width=0.41\textwidth]{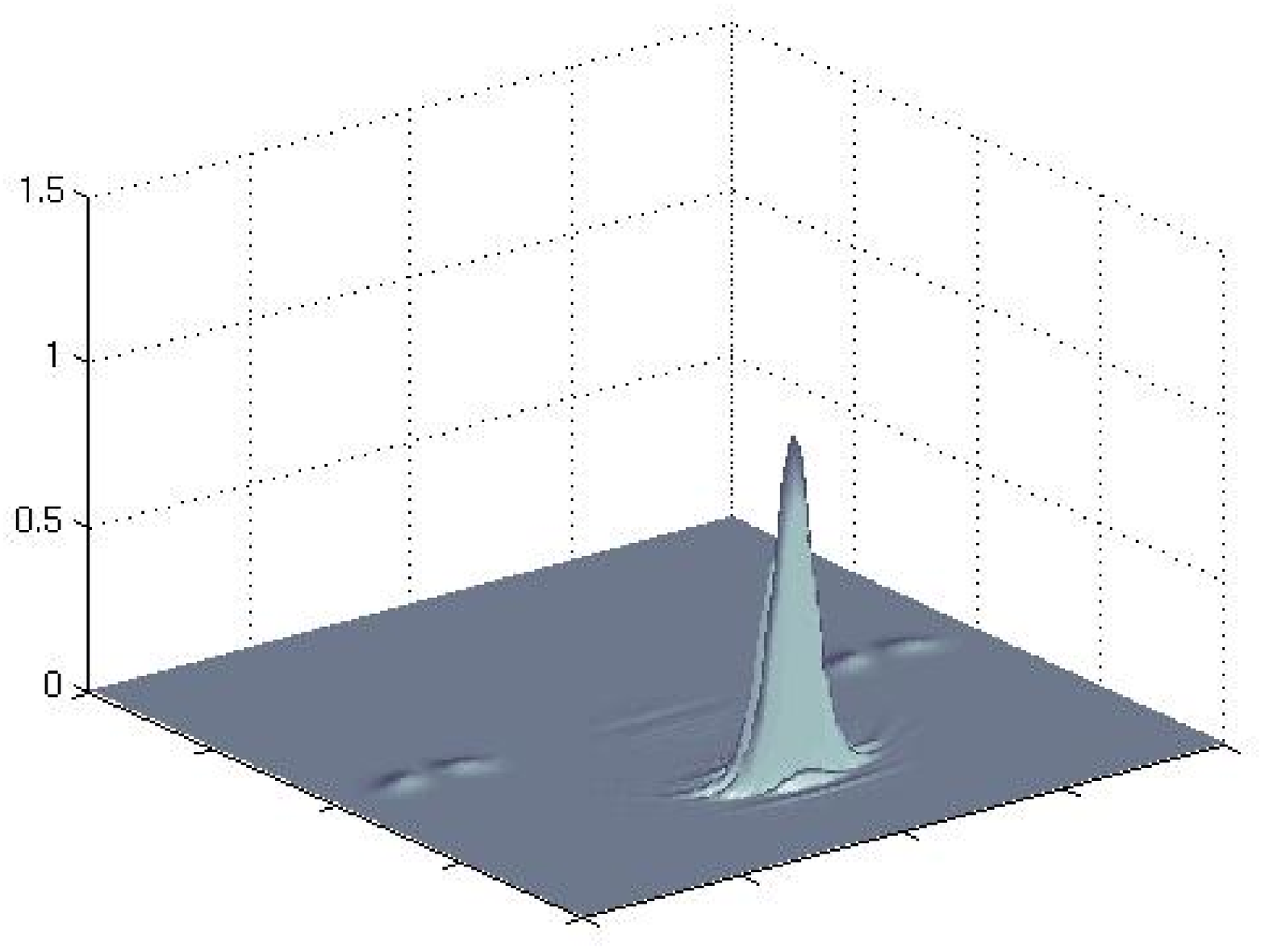}
\includegraphics[width=0.41\textwidth]{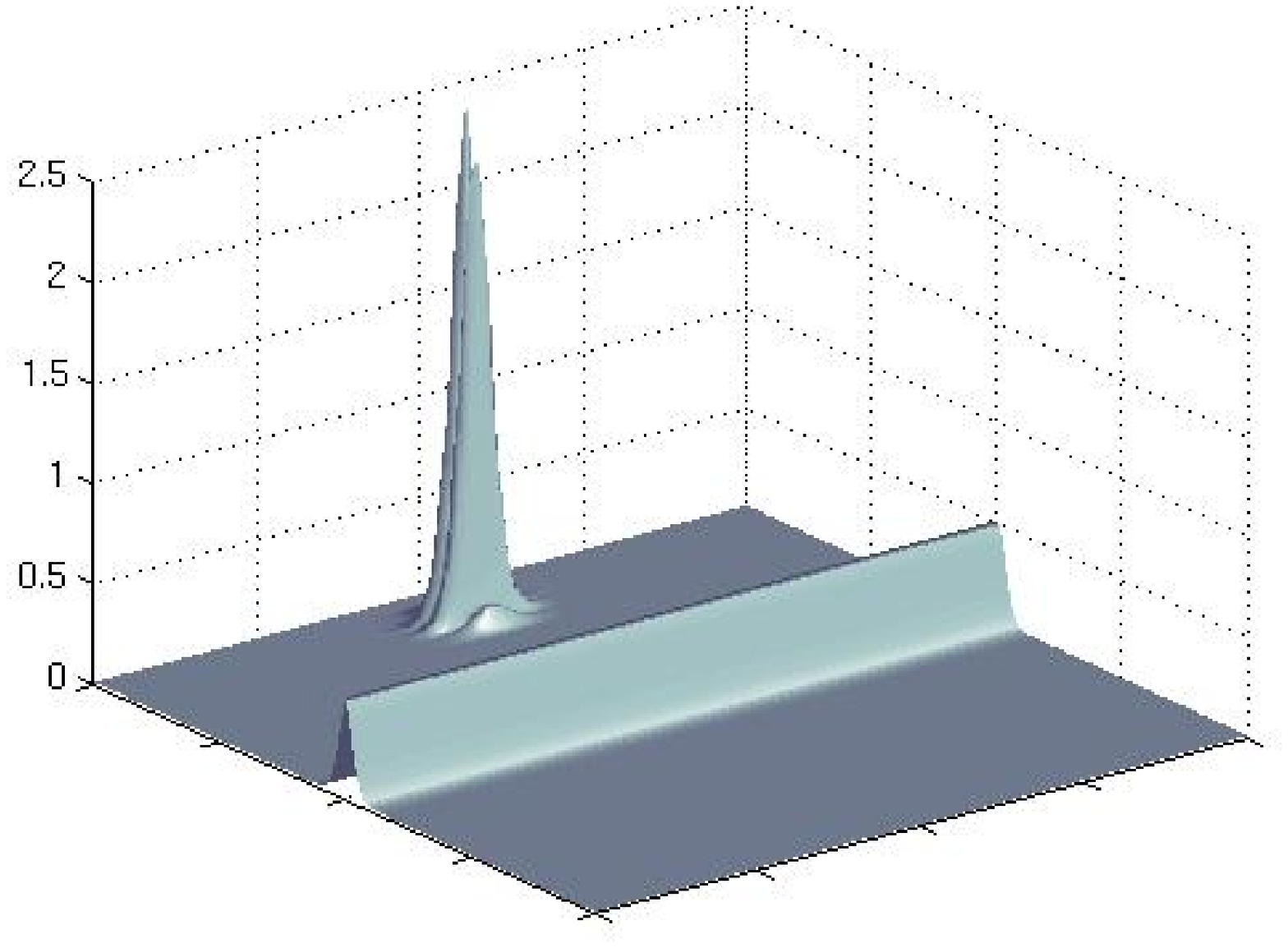}
\includegraphics[width=0.41\textwidth]{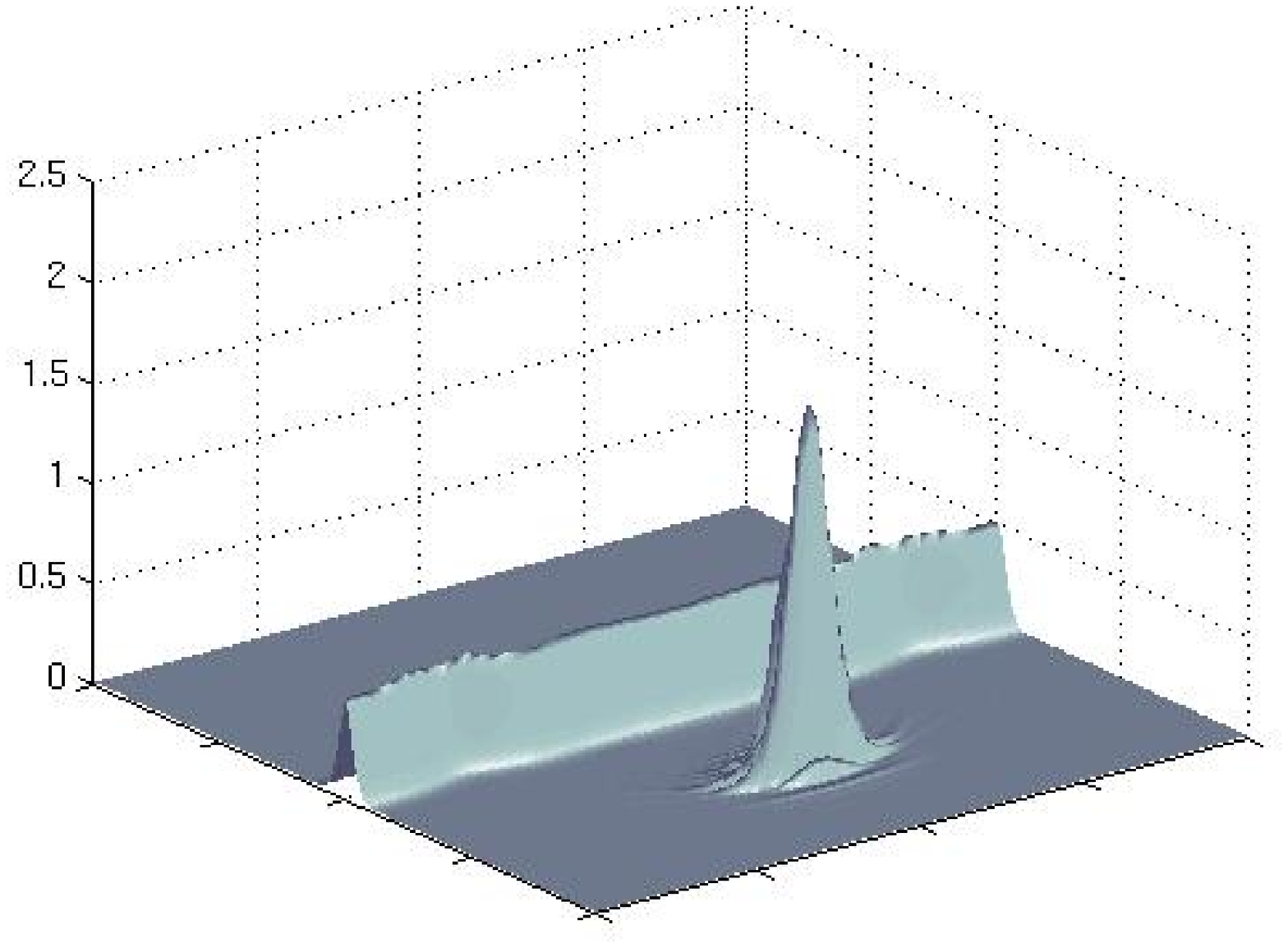}

  \caption{\label{f:oscillon_domainwall} An oscillon crossing a domain wall: 
upper panel show the value of field before (left) and after (right) the 
collision, 
middle shows the kinetic energy density and total energy below.}

\end{figure*}

As oscillons are made by domain walls and survive collisions, 
it is natural to pose the question what happens when an oscillon 
meets a domain wall.
Figure~\ref{f:oscillon_domainwall} shows snapshots of an oscillon with 
an initial velocity $v \simeq 0.75$ crossing a domain wall. 
Oscillon is clearly recognisable both 
before the encounter and afterwards oscillating around the other vacuum.
Snapshots of the total energy density show that crossing has caused a 
perturbation on the domain wall that propagates at the speed of light 
along the wall away from the interaction point.
Snapshots in the kinetic energy where the static domain wall is initially 
invisible show that while oscillon has shed some energy to the domain wall, 
that is a relatively tiny fraction as the ripples along the domain wall 
are barely visible. Though the direct energy transfer between the oscillon 
and the domain wall is in this case relatively small, oscillon is deformed, 
slightly elongated in the direction of the domain wall, and potentially 
radiates some its energy.
It should be emphasised that the presented encounter is not necessarily 
a typical one. 
Apart from the velocity, the relative phase of the 
oscillon seems to strongly control the amount of energy transfer 
in a collision. 
{There are two effects that readily seem to enhance the crossing 
probability at higher velocities. Lorentz contraction shortens the length 
of the disturbace oscillons creates and the lower 
frequency and thus slower time-evolution of the wave decreases 
the energy transfer to the domain wall.}
In any case, just the potential of oscillons crossing 
domain walls and propagate from one vacuum to another demonstrates 
how surprisingly persistent objects they are.

\section{Conclusions}

We have studied the field dynamics of a quartic double well 
potential with random initial conditions in two spatial dimensions. 
We have shown good 
evidence that when the field is undamped the collapse of domains takes 
place so rapidly that there is enough energy to excite the field into 
long-lived, non-perturbative oscillating energy concentrations.
Furthermore, we have examined oscillons in a less radiative 
environment but in still a random system and shown that a fraction of 
oscillons persist for a long time. We reported on merging of oscillons 
in off-axis collisions at low velocities as well as the potential of 
oscillons to cross a domain wall at high enough velocity.
Unfortunately the method inspired by the spectral function 
does not provide unambiguous information due to the suppression 
of signal with increasing velocity. On the other hand long 
domain walls leave a very distinctive trace and maybe similar techniques 
could be used in studies of defects as well.

We did not impose any damping for the system apart from at very early 
times to condense it. 
Damping is not likely to alter the process of domain collapse 
considerably, unless the dissipation is really strong, while the 
further evolution may be different. However, it is not clear what kind of 
consequencies dissipation would have. 
Damping could reduce the dispersive radiation modes and thus even 
enhance the oscillon lifetime. 
Obvious reason for friction in the system is the Hubble damping in the 
Early Universe. A study in one dimension showed that oscillons could 
persist in an expanding background when the expansion rate is low 
enough~\cite{Graham:2006xs}. 
Expansion can reduce 
velocities which may once again have twofold consequencies, oscillon may 
merge and demise in collision easier, but may also reduce collision 
rate and then increase the lifetime.


\begin{acknowledgments}

P.S. thanks Gert Aarts, Szabolcs Borsanyi, 
Ajit Srivastava, Anders Tranberg and Tanmay Vachaspati 
for discussions.
P. S. was supported by 
the Netherlands Organization for Scientific Research (N.W.O.) 
under the VICI programme. 
This work was in part initiated when P.S. was supported by
Marie Curie Fellowship of the
European Community Program HUMAN POTENTIAL 
under contract HPMT-CT-2000-00096.
 
\end{acknowledgments}

\appendix

\section{}

We present here an analytic outline how moving Gaussian distribution 
would appear in spectral function $\rho(\omega)$. 
The starting point are the 
ans\"atze~(\ref{gaussian-ansatz}) and~(\ref{solution-ansatz}). 
Assume now further that all the modes $f_{n}(r)$ have the Gaussian form 
with the same width
\begin{eqnarray}
\phi(r,t) = \exp (-r^2/r_{0}^2) \cdot \sum_{n=1}^{\infty} 
a_{n} \, \cos(n \omega_{0} t) \,.
 \label{appendix1}
\end{eqnarray}
Consider now an arbitrary term in the sum under a boost with velocity $v$ 
in $x_{1}$-direction. This yields
\begin{eqnarray}
a_{n}
\exp \Big(- \frac{\gamma^2 u^2 + x_{2}^2}{r_{0}^2} \, \Big)
\cdot 
\cos \Big( n \omega_{0} \big( \frac{t}{\gamma} -\gamma v u \big) \Big) \, ,
\label{appendix2}
\end{eqnarray}
where we have defined $u = x_{1} - vt$.
The volume average integrating over the variables $u$ and $x_1$ yields
\begin{eqnarray}
\frac{a_{n}}{\gamma} \,
\exp \Big(- \frac{n^2 \omega_{0}^2 r_{0}^2}{4} \, v^2 \Big) \cdot
\cos \Big( \frac{n \omega_{0} t }{\gamma} \Big) \, .
\label{appendix3}
\end{eqnarray}
The decrease in the freqiency is now explicitly readable in the 
time dependent part. Turning to the basic frequency, set $n=1$ 
(it is also noticible that the higher modes $n>1$ that appear with weaker 
amplitudes $a_{n}$ undergo an extra suppression, which is to a certain extent 
visible in Figure~\ref{f:spectral}, compared to the static situation $v=0$).
Spectral function in the approximation used is the 
product $\bar{\Pi} \cdot \bar{\phi}$. Thus its amplitude has functional 
dependence on the kinematic variables $v$ and $\gamma$
\begin{eqnarray}
A(v) \propto
\frac{1}{\gamma^3} \,
\exp \Big(- \frac{1}{2} \, \omega_{0}^2 \, r_{0}^2 \, v^2 \Big) \, .
\label{appendix4}
\end{eqnarray}
Consider now an arbitrary velocity distribution $n(v)$. 
This yields a signal in spectral function
\begin{eqnarray}
\rho(\omega) \propto \int dv \, n(v) \, A(v) \, 
\delta (\omega - \omega _{0}/\gamma(v)) \, .
\label{appendix5}
\end{eqnarray}
By the simple relation $\gamma=\omega_{0} / \omega$ this can be 
turned to the velocity distribution
\begin{eqnarray}
n(v) = \frac{\omega_{0}\sqrt{\omega_{0}^2-\omega^2}}{\omega} \,
\frac{\rho(\omega)}{A(\omega)} \, .
\label{appendix6}
\end{eqnarray}


\bibliography{oscillon}
\include{oscillon}

\end{document}